\documentclass[acmsmall,authorversion,nonacm]{acmart}

\usepackage{tabularx}%
\setcopyright{none}
\copyrightyear{2022}
\acmYear{2022}

\makeatletter
\let\@authorsaddresses\@empty
\makeatother

\begin{document}

\title[Taxonomy of Prompt Modifiers]{A Taxonomy of Prompt Modifiers for Text-To-Image Generation}%

\author{Jonas Oppenlaender}
\email{joppenlu@jyu.fi}
\affiliation{%
  \institution{%
  University of Jyv\"askyl\"a}
  \country{Finland}
  \postcode{40014}
}


\begin{abstract}
Text-guided synthesis of images has become enormously popular and online communities dedicated to text-to-image generation and art generated with Artificial Intelligence (AI) have emerged.
While deep generative models can synthesize high-quality images and artworks from simple descriptive text prompts, practitioners of text-to-image generation typically seek to control the generative model's output by adding short key phrases (``modifiers'') to the prompt.
This paper identifies six types of prompt modifiers used by practitioners in the online text-to-image community based on a 3-month ethnographic study. The novel taxonomy of prompt modifiers provides researchers a conceptual starting point for investigating the practice of text-to-image generation, but may also help practitioners
of AI generated art improve their images.
We further outline how prompt modifiers are applied in the practice of ``prompt engineering.''
We discuss research opportunities of this novel creative practice in the field of Human-Computer Interaction (HCI). The paper concludes with a discussion of broader implications of prompt engineering from the perspective of Human-AI Interaction (HAI) in future applications beyond the use case of text-to-image generation and AI generated art.






\end{abstract}




\maketitle

\section{Introduction}

Text-to-image generation has become widely popular both in academia and as a new creative practice among practitioners of ``AI art.''
Based on deep learning, text-to-image generation systems can generate digital images from short descriptive texts (called \textit{prompts}, such as \textit{``an oil painting of a beautiful landscape at dawn''}).
To be effective, the textual input prompts need to be given in a certain format in order to, for instance, generate images with a certain style.
This is commonly achieved by adding keywords and key phrases to the prompt (so-called ``prompt modifiers'').
Examples of images synthesized from textual prompts are depicted in \autoref{fig:examples}.
Given the quality of these images, it is not surprising that an enthusiastic online community around this novel text-based way of creating images and art has developed. Within this community, the practice and skill of writing prompts is known by the term ``prompt engineering'' due to its iterative and experimental nature~\citep{chilton}.
Prompt engineering is an emerging research area in the field of Human-Computer Interaction~(HCI) concerned with how to phrase input prompts for deep generative models and -- from a broader perspective -- how humans can effectively interact with  artificial intelligence.


The learning curve of prompt engineering can still be steep. Some prompt modifiers used within the community of practitioners are not intuitive and 
from looking at an image, it is impossible to tell the input prompt used to synthesize the image.
On social media, many artists do not share their complete prompts for their artworks and it is often not clear how these artworks were created.
Therefore, prompt engineering is a non-intuitive skill that is learned from extensive experimentation and trial and error~\citep{chilton}.
A growing number of resources in the gray and scholarly literature present systematic experimentation on the effect of different prompt modifiers~\citep{chilton,artstudies,DiscoDiffusionArtiststudies,dallepromptbook,travelersguide}.
Online databases have been created in which users can explore artworks, prompts, and prompt modifiers \citep[e.g.][]{2210.14896.pdf,2022.acl-demo.9.pdf,arthub.ai,lexica.art,openArt.ai}.
These resources and guides are part of a growing online ecosystem around text-to-image generation \citep{creativitypromptengineering}.

\newlength{\thwidth}%
\setlength{\thwidth}{4.3cm}%
\begin{figure}[!ht]
\begin{center}
\begin{tabular}{rcl}%
    \includegraphics[width=\thwidth]{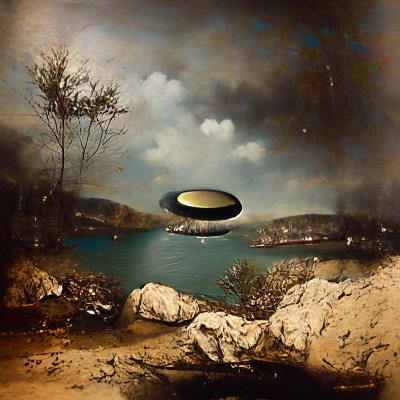}%
&
    \includegraphics[width=\thwidth]{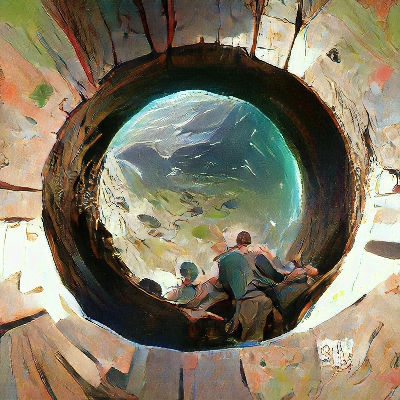}%
&
    \includegraphics[width=\thwidth]{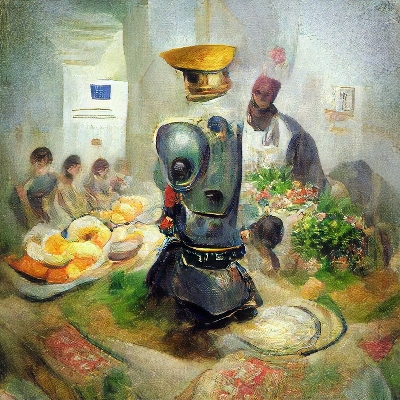}%
\\[.2cm]
    \includegraphics[width=\thwidth]{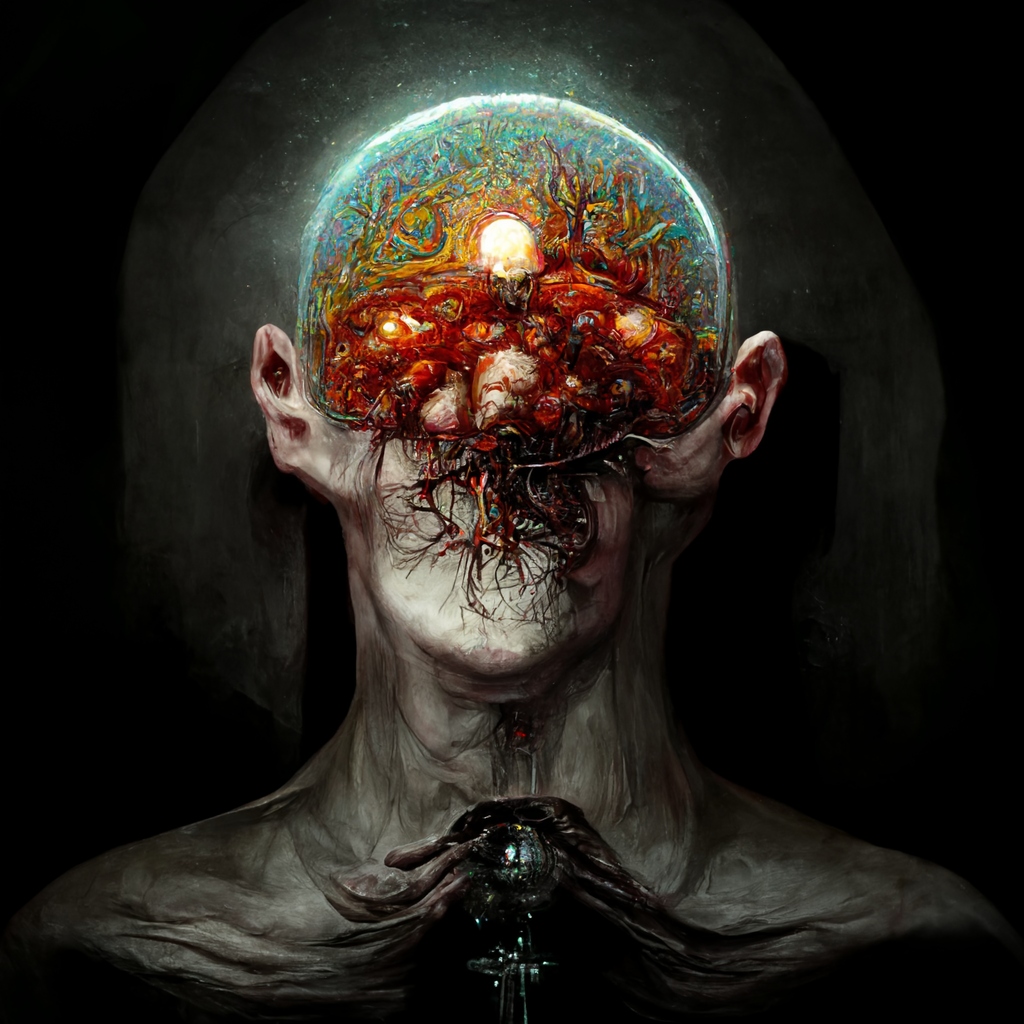}%
&
    \includegraphics[width=\thwidth]{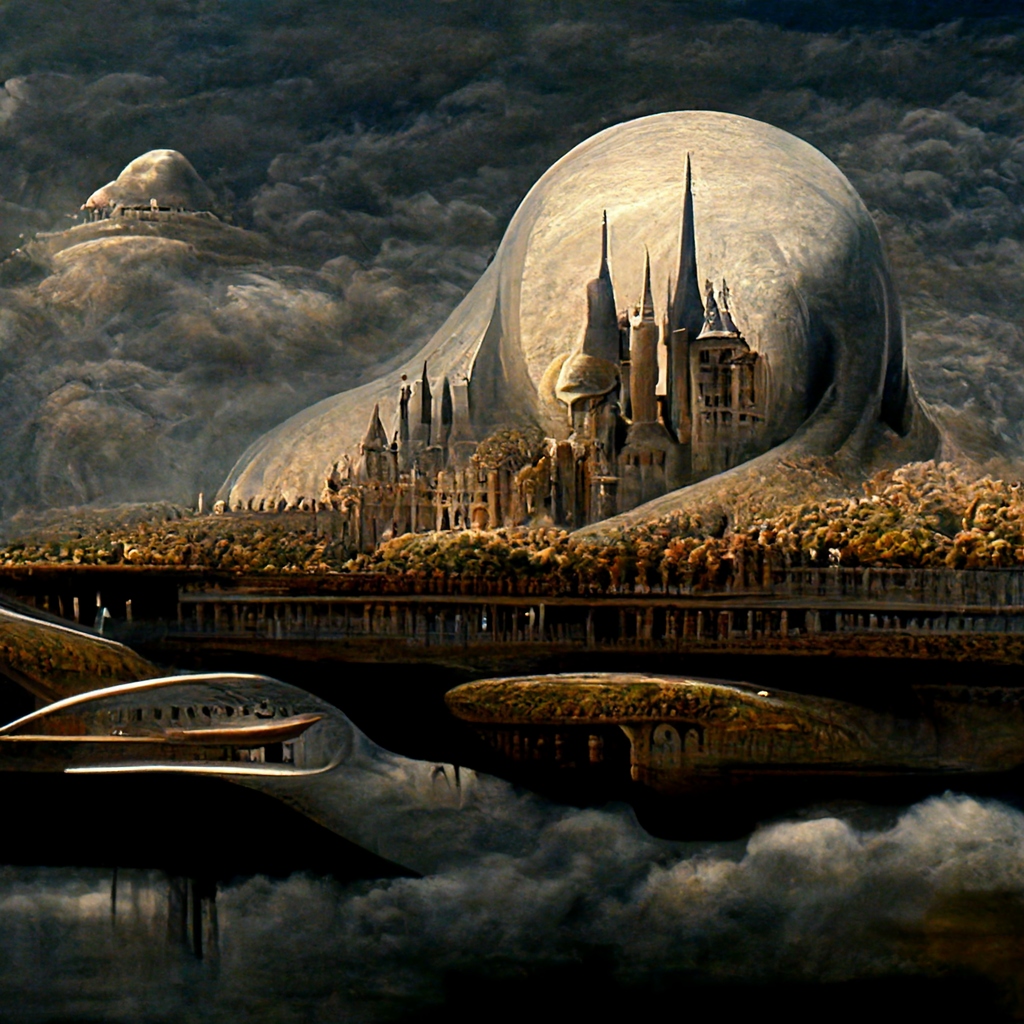}%
&
    \includegraphics[width=\thwidth]{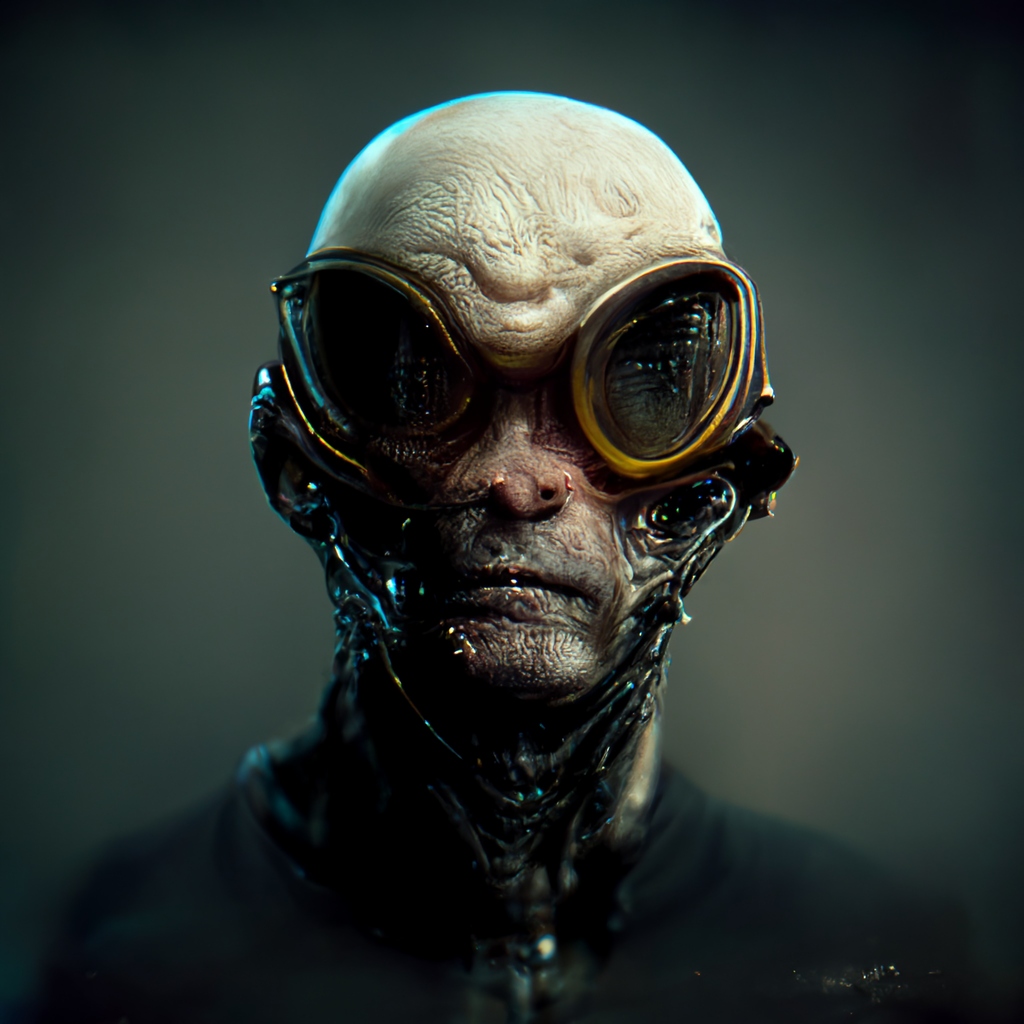}%
\\[.2cm]
    \includegraphics[width=\thwidth]{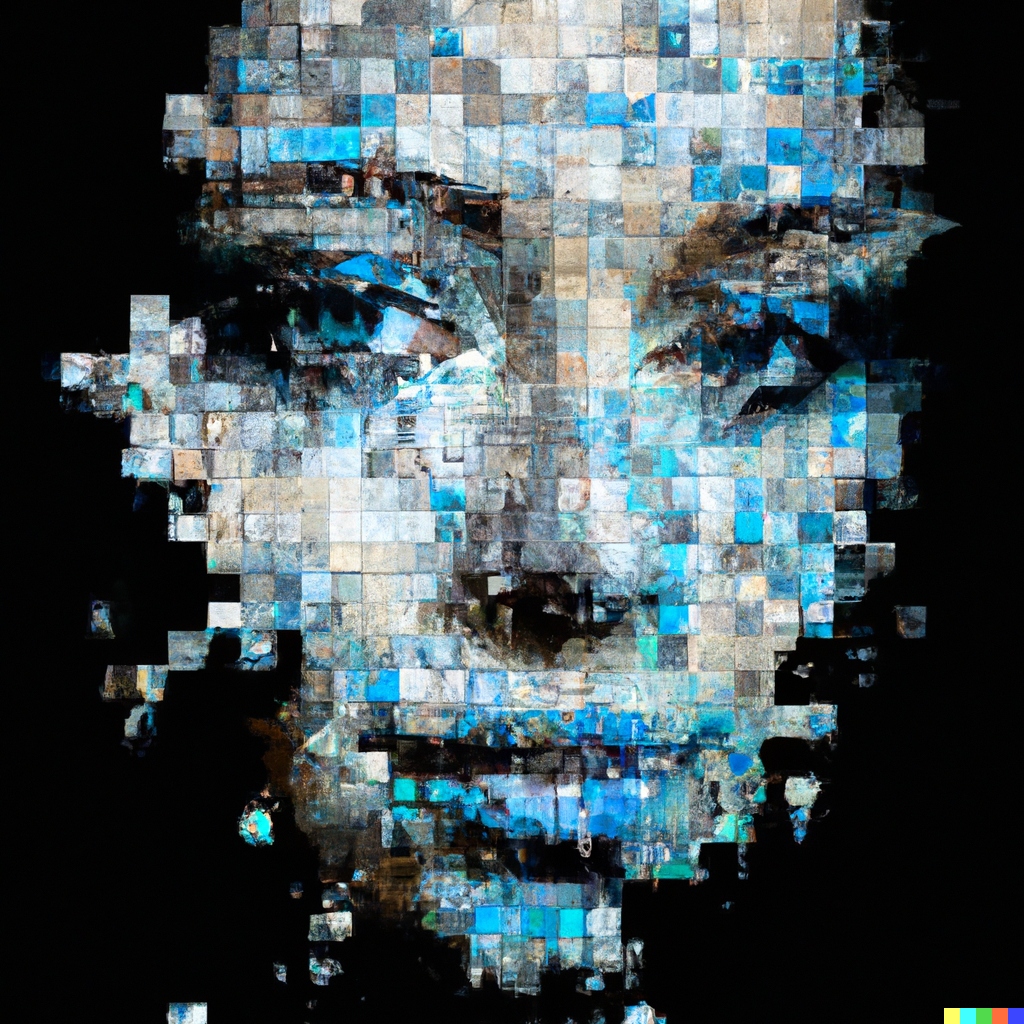}%
&
    \includegraphics[width=\thwidth]{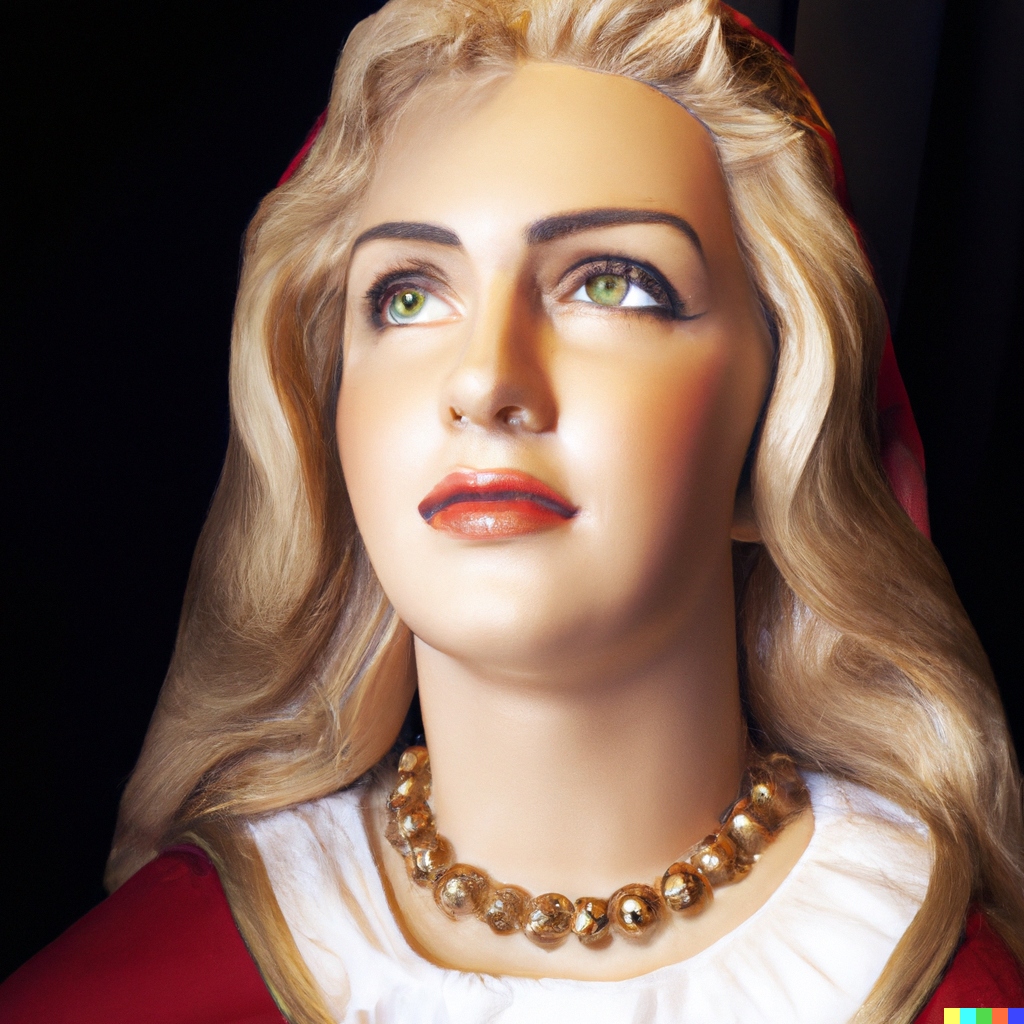}%
&
    \includegraphics[width=\thwidth]{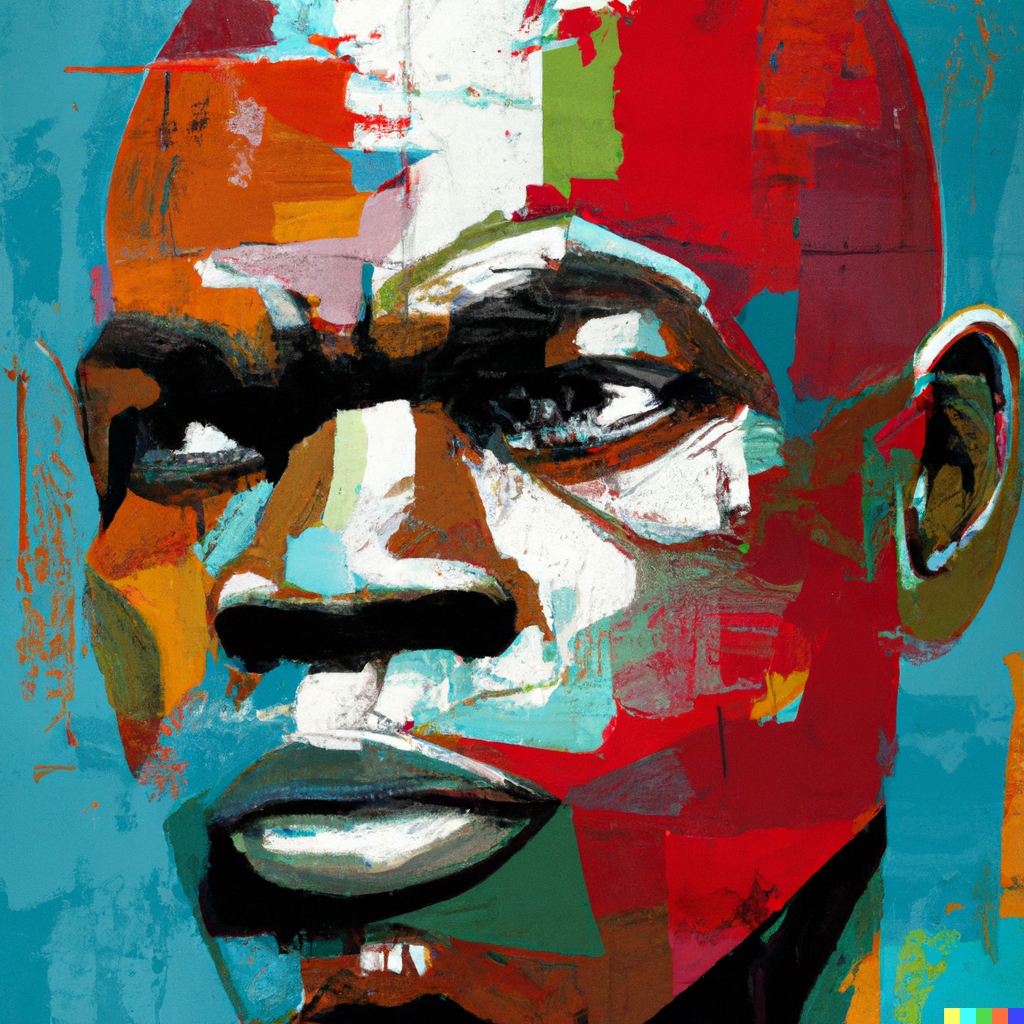}%
\end{tabular}%
\end{center}
\caption{Selected images generated with text-to-image generation using VQGAN--CLIP (top), 
Midjourney.com (middle),
and DALL-E~2 (bottom).}
\label{fig:examples}%
\end{figure}

While guides, resources, and datasets about prompting are available, there is still a gap in our understanding of prompt modifiers.
No previous study has investigated different types of prompt modifiers.
With a specific focus on digital art generated with text-to-image systems, this paper contributes a taxonomy of prompt modifiers used by practitioners in the text-to-image community, based on an 
ethnographic study of the community's prompt engineering practices. 
The work is based on a three-month online ethnography which analyzed how prompt modifiers are being applied in prompt writing. 
This paper contributes toward a better understanding of prompt engineering as a practice within HCI in order to inform the HCI research community on the emerging practice of prompt engineering within the broader context of human interactions with artificial intelligence.
This paper aims to enhance the theoretical understanding of how people write prompts and use prompts modifiers.
Through understanding prompt writing, we can pave the way towards a broader and unified theory of prompt engineering which the HCI literature is currently missing.
The paper also touches on how the technology behind text-to-image systems and the practice of prompt engineering has broader implications in research on HCI and Human-centered AI (HCAI).

The paper is structured as follows.
We first provide a brief introduction into text-to-image synthesis and prompt engineering in Section~\ref{sec:background}.
After describing the methodological approach in Section~\ref{sec:method},
a taxonomy of six different types of prompt modifiers 
is presented in Section~\ref{sec:modifiers}.
It is demonstrated how these prompt modifiers are applied in the context of prompt engineering in Section \ref{sec:promptengineering}.
The paper concludes with a discussion of opportunities for future research on text-to-image generation and the broader implications beyond AI generated art (sections~\ref{sec:discussion} and \ref{sec:conclusion}).

\section{Background}%
\label{sec:background}%
%
This section discusses the evolution of text-to-image generation, particularly highlighting the role of OpenAI’s CLIP model. It then delves into the concepts of ``prompt engineering,'' a creative practice for controlling image generation, and ``prompt modifiers,'' keywords used to refine the image output.
The section also underscores the contribution of the online community in advancing these creative practices.

\subsection{Text-to-Image Generation}%
The field of image synthesis using deep learning has seen an unprecedented growth with the break-through development of multimodal models trained on large amounts pairs of of images and text scraped from the World Wide Web.
%
The development was initially spurred by OpenAI's multimodal model CLIP~\citep{CLIP}.
    CLIP is a contrastive language-vision model trained in an unsupervised way to perform zero-shot classification of images.
    CLIP provides a convenient way to transform both text and images into a common vector-based representation.
    When used as a discriminator component in text-conditioned generative systems, CLIP can ``guide'' the image generation process.
CLIP was originally a part of OpenAI's DALL-E architecture \citep{2102.12092.pdf}, a text-to-image system that was never released in its entirety.
However, OpenAI did release the weights of the CLIP model.
This resulted in a vast number of open source implementations of text-to-image systems, first as CLIP-guided generative adversarial networks (e.g., {VQ\-GAN}--{CLIP} by \citet{VQGANCLIP}) and later as diffusion based image generation systems, such as CLIP Guided Diffusion~\citep{CLIPguideddiffusion} and Latent Diffusion~\citep{latent-diffusion}.

This paper investigates text-to-image generation from the lens of Human-Computer Interaction (HCI).
    In order to generate images from text, one not only has to choose the right words to make the text-to-image system generate the desired images, one also has to add different keywords and key phrases to control the style and quality of the image generation.
This creative practice 
of writing effective prompts is sometimes referred to as ``prompt engineering.''
    This paper investigates what (and how) different types of prompt modifiers are being applied in prompt engineering.


\subsection{Prompt Engineering and Prompt Modifiers}%
\label{sec:text-prompts}%
Prompt engineering~\citep{chilton}
    -- also referred to as ``prompt design''~\citep{openaidocs}, ``prompt programming''~\citep{2102.07350.pdf}, and ``prompting'' \citep{2022.acl-demo.9.pdf} for short --
is the practice of writing textual inputs for generative systems.
In the context of text-to-image generation, ``carefully selected and composed sentences are used to achieve a certain visual style in the synthesized image''~\citep{2207.13038.pdf}.
The practice has seen an ideal application ground in AI generated art, but it is not limited to text-to-image generation.
The term prompt engineering was originally coined to denote the practice of writing textual inputs for the language model GPT-3~\citep{chilton}.
    This autoregressive language model requires context to produce relevant text as output. Templates have been developed to optimally provide textual inputs to GPT-3. OpenAI's documentation, for instance, lists 49 ``recipes'' on how to phrase input prompts for their language model.\footnote{See https://beta.openai.com/examples.}
Templating languages and interfaces have been developed to advance the field of prompting \citep{2022.acl-demo.9.pdf}.
    Using such recipes and tools, the output of the language model can be adapted to different down-stream tasks, such as correcting grammar, summarizing text, answering questions, generating product names, or acting as a chat bot.

\begin{figure}[h!]
\begin{center}
  \includegraphics[width=.8\linewidth]{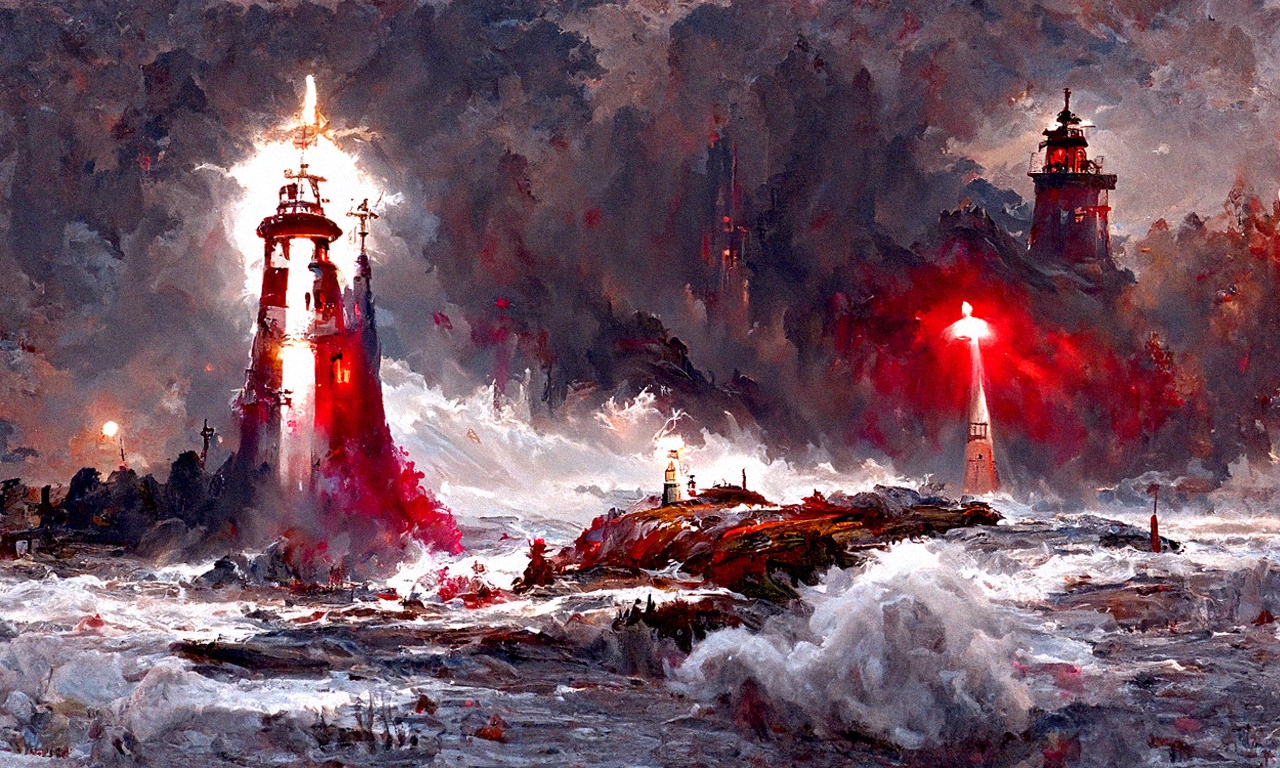}%
\end{center}
\caption{Digital artwork generated with DISCO Diffusion from the input prompt \textit{``A beautiful painting of a singular lighthouse, shining its light across a tumultuous sea of blood by greg rutkowski and thomas kinkade, Trending on artstation.''} This prompt is part of the default configuration settings in the DISCO Diffusion notebook.{\protect\footnotemark}}%
\label{fig:lighthouse}%
\end{figure}

Templates 
have also emerged for writing input prompts for text-to-image systems, particularly in the online community around AI generated art.
    For instance, the ``Traveler's Guide to the Latent Space'' recommends the following prompt template~\citep{travelersguide}:
\begin{quote}
    \textit{[Medium][Subject][Artist(s)][Details][Image repository support]}
\end{quote}
Similar templates are being followed in many resources originating from within the online community, such as the DALL-E Prompt Book~\citep{dallepromptbook}.
\autoref{fig:lighthouse} provides an example of a typical textual input prompt and the resulting AI generated image.

\footnotetext{See https://github.com/alembics/disco-diffusion.}

Prompt engineering is not a hard science as found in the fields of science, technology, engineering, and mathematics (STEM). Rather, it is a term that originates from within the online community of practitioners of text-to-image generation. The term reflects the community's self-understanding, similar to the terms ``AI art'' and ``AI artist'' which also originate from within the community.
Due to the rise in popularity of text-to-image systems, practitioners of AI art include not only technology-savvy developers and early-adopting hobbyists,  but also artists, professionals, semi-professionals, and ``Pro-Ams''~\citep{2556288.2557298.pdf} with or without commercial interests.
In the remainder of this paper, we will refer to the members of the online text-to-image community as \textit{practitioners}.

Prompt engineering resembles a conversation with the text-to-image system. 
A practitioner typically will run a prompt, observe the outcome, and adapt the prompt to improve the outcome.
Prompt engineering, thus, is iterative and practitioners formulate prompts as probes into the generative models' latent space.
The online community quickly found that the aesthetic qualities and subjective attractiveness of images can be improved by adding certain keywords and key phrases to the textual input prompts. The terms may be referred to by a number of different names, such as ``style phrases,'' ``clarifying keywords'' \citep{2209.11711.pdf}, or ``vitamin phrases''~\citep{pressmancrowson2022}.
In this paper, we refer to them as \textit{prompt modifiers}. By adding a prompt modifier to a textual input, one seeks to direct the text-to-image system in certain directions, hence ``modifying'' the resulting image.

\begin{figure}[!ht]
\begin{center}
  \begin{tabularx}{\textwidth}{XXXX}%
        \includegraphics[width=\linewidth]{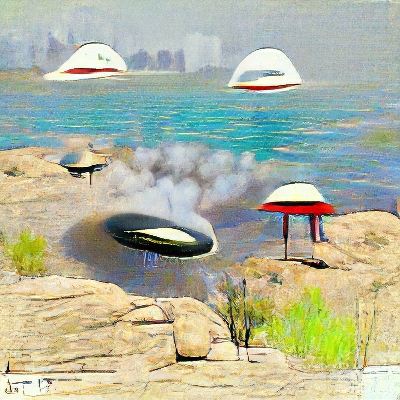} &
        \includegraphics[width=\linewidth]{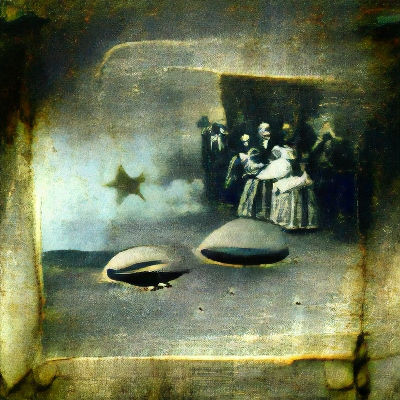} &
        \includegraphics[width=\linewidth]{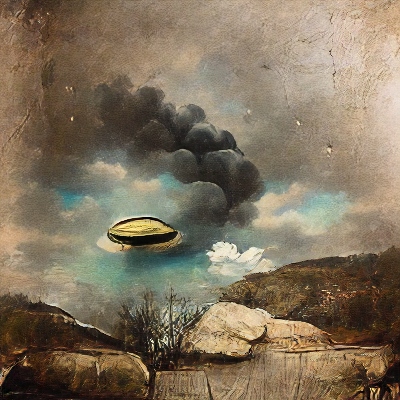} &
        \includegraphics[width=\linewidth]{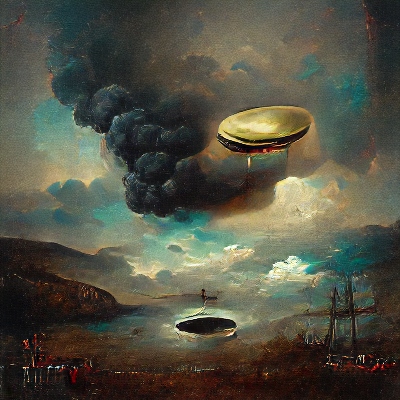}
    \\%
            \multicolumn{1}{c}{a)}
        &
            \multicolumn{1}{c}{b)}
        &
            \multicolumn{1}{c}{c)}
        &
            \multicolumn{1}{c}{d)}
    \\%
    \multicolumn{4}{l}{
        \small
            Text prompts:
    } \\
    \multicolumn{4}{l}{
        \small
            a) \textit{``ufo landing''}
    } \\
    \multicolumn{4}{l}{
        \small
            b) \textit{``ufo landing, daguerreotype''}
    } \\
    \multicolumn{4}{l}{
        \small
            c) \textit{``ufo landing, daguerreotype, trending on /r/art''}
    }\\
    \multicolumn{4}{l}{
        \small
            d) \textit{``ufo landing, daguerreotype, by greg rutkowski, trending on /r/art''}
    }
  \end{tabularx}%
\end{center}
\caption{Example of iterative prompt engineering for generating an image. Images generated with VQGAN--CLIP by \citet{VQGANCLIP} with
    175 iterations,
    CLIP model ViT-B/32,
    VQGAN model wikiart\_16384,
    and
    seed 6087304447281500163.}%
\label{fig:promptengineering}
\end{figure}

In practice, prompt modifiers are applied through experimentation or based on best practices learned from experience or online resources.
An example of an iterative application of prompt modifiers can be seen in Figure~\ref{fig:promptengineering}.
Knowing what prompt modifiers work best for a given subject term is often the result of the practitioner's iterative experimentation, research in online communities, and the use of online tools and resources created for supporting the practice of prompt engineering~\citep{creativitypromptengineering}.

\section{Method}%
\label{sec:method}%
%
%
In this research, a dual-methodological approach was adopted, leveraging both autoethnographic (Section~\ref{sec:autoethnography}) and online ethnographic studies (Section~\ref{sec:ethnography}), to delve into the nuanced aspects of prompt engineering and text-to-image art generation. Understanding the intricacies of prompt engineering, an acquired skill cultivated through iterative experimentation, necessitates a hands-on, experiential approach. Hence, autoethnography provided a fitting method to gain an intimate, practitioner’s perspective. By conducting an autoethnographic study, the author was able to engage with the process of text-to-image synthesis personally, thereby capturing its nuances from a first-hand perspective. However, the complex and communal nature of this emerging field necessitated a broader perspective. To capture the collective wisdom and shared practices within the field, the author further complemented the autoethnographic approach with an online ethnography of the text-to-image art community. This approach allowed the author to glean insights from the shared experiences and resources of the broader community of practitioners active in online spaces, primarily on Twitter. The synthesis of these two complementary approaches aimed to provide a comprehensive understanding of prompt engineering, bridging the gap between individual experience and collective knowledge.




\subsection{Autoethnographic Research on Prompt Engineering}
\label{sec:autoethnography}%
%
Prompt engineering is learned through iterative experimentation akin to ``brute-force trial and error''~\citep{chilton}.
Therefore, prompt engineering is an acquired skill that is associated with a learning curve.
The skill can be learned from community-provided resources, such as written guides and reports of systematic experimentation, or from prompts shared on social media, such as online communities dedicated to text-to-image art \citep{creativitypromptengineering}.
However, to appreciate and understand the craft of prompt engineering and text-to-image generation, one has to apply the knowledge and experiment with different input prompts.
Autoethnography research is, therefore, an appropriate method to learn about prompt engineering.

The author conducted a 3-month autoethnographic study~\citep{10.2307_23032294.pdf,handbook,160940690400300403.pdf}
between October 2021 and December 2021.
This personal ethnography~\citep{crawford1996.pdf} allowed the author to  get a ``practitioner’s perspective''~\citep{160940690400300403.pdf} of text-to-image generation by ``learning from self-use''~\citep{p514-neustaedter.pdf}.
The author experimented with text-to-image synthesis and
created digital images with a text-to-image system using notebooks hosted on Google's Colaboratory (Colab).\footnote{https://colab.research.google.com}
The author started on average at least one Colab session every work day between October 4 and December 31, 2021.
The free tier of Google Colab was used in all sessions. This limited the overall working time to about 2 hours per day, depending on the computational power of the assigned resources and whether penalties were incurred the previous day.
%
VQGAN--CLIP by \citet{VQGANCLIP} was chosen as text-to-image system using a notebook titled ``VQGAN and CLIP (z + quantize method with augmentations)''.\footnote{https://colab.research.google.com/github/justinjohn0306/VQGAN-CLIP/blob/main/VQGAN\%2BCLIP(Updated).ipynb} This VQGAN--CLIP notebook was originally created by Katherine Crowson, with 
``modifications by Eleiber \# 8347'' and a ``friendly interface'' by ``Abulafia \# 3734'' and further modifications by Justin John.
VQGAN--CLIP was selected for several reasons.
    First, VQGAN--CLIP was one of the first text-to-image systems that experienced widespread popularity in the emerging text-to-image art community in 2021. This made VQGAN--CLIP instrumental to the growth of the community~\citep{VQGANCLIP}.
    Second, the system can be executed on Google's Colaboratory (Colab) free of charge. The system requires less memory than later systems, and it is therefore less likely that image generation will fail due to insufficient memory.
    Third, the VQGAN--CLIP notebook on Colab is very accessible and straight-forward to use, with only a small number of configuration parameters (cf. \autoref{fig:promptengineering}).
    Last, the system is deterministic. Consecutive runs with the same configuration parameters will produce exactly the same images which makes the images reproducible. This is not the case with some of the later systems which make use of non-deterministic algorithms.
The author generated 885 images in the course of this study.

The autoethnographic research was not conducted from scratch. Rather, it was informed by learning from the community on social media. To this end, the autoethnographic research was complemented with an online ethnography of the text-to-image art community on Twitter and a study of online community resources, described in the following section.

\subsection{Ethnographic Study of the Text-to-Image Art Community}%
\label{sec:ethnography}%
An ethnographic study of prompt engineering
was conducted on 
Twitter (see Section \ref{sec:twittercommunity}). 
The aim of this social media ethnography~\citep{postill2012.pdf,pink2016} was to learn more about the textual prompts used in the community of practitioners of text-to-image art.
Insights derived from the study of this community were used in the autoethnographic experimentation with the text-to-image system.
The research was complemented with a review of the literature (Section \ref{sec:literaturereview}).

\subsubsection{Twitter community}%
\label{sec:twittercommunity}%
A dedicated online community around text-to-image generation with specific focus on AI generated art -- which practitioners in the online community sometimes refer to as ``AI art''~\citep{aiart} --
has emerged.
Social media services, such as Twitter, are a well-suited outlet for practitioners in this community to post and share images and experiences.

During the 3-month period of research, the author took the role of ``participant-as-observer''~\citep{2573808.pdf} by engaging with the text-to-image art community  on Twitter, participating in discussions, and posting images created with the text-to-image system.
The author followed posts on Twitter to learn about different prompts used in the text-to-image art community.
To this end, the author followed trending hashtags, such as
	\#vq\-gan\-clip,
	\#VQ\-GAN,
	\#clip\-guided\-diffusion,
	\#digital\-art,
	\#AI\-Art, and
	\#generative\-art.
%
Not every practitioner of text-to-image art shares their prompts on Twitter. 
Especially if commercial interests are involved~-- e.g., selling the art as non-fungible tokens (NFTs)~-- practitioners may keep their prompts a secret.
The research material, therefore, was sparse at the time of conducting the study.
However, some practitioners are more liberal in sharing their prompts. It is the posts from this group of Twitter users that informed this research (e.g., posts by Katherine Crowson (@RiversHaveWings), Hannah Johnston (@hannahjdotca), @nshepperd1, and John David Pressman (@jd\_pressman), to name but a few).


\subsubsection{Review of community resources}%
\label{sec:literaturereview}%
In parallel to the research on the online community,
a review of the literature was conducted, with specific focus on text-to-image generation and the practice of prompt engineering for digital art.
With the exception of \citeauthor{chilton}'s design guidelines for prompt engineering~\citep{chilton,3527927.3532792.pdf}, there still is little scholarly literature on the practice of text-to-image generation for AI generated art in the field of HCI.
Therefore, the literature review primarily focused on sources in the gray literature, such as community-provided resources, documents, guides, experiment reports, blog posts, articles on the Web.

\subsection{Inductive Development of the Taxonomy}%
\label{sec:taxonomymethod}%
The taxonomy was developed inductively from pieces of information found during the research.
Due to the relative scarcity of this material at the time of writing, the development of the taxonomy was conducted iteratively, as follows.
%
%
%
A list of potential candidates for prompt modifiers was inductively compiled and 
grouped.
This list was subject to continual reinterpretation when novel instances of prompts were encountered.
	Whenever a candidate for a novel type of prompt modifier was found in a post on Twitter or the literature, the author revisited the list of prompt modifiers.
	Therefore, the resulting taxonomy was iteratively and inductively revised and expanded when new types of prompt modifiers were encountered.
After some weeks of collecting data this way, the list of prompt modifiers and taxonomy did no longer grow, even if instances of novel and atypical prompts were encountered.
This indicates the completeness of the developed taxonomy.

The findings were documented in a PowerPoint presentation
with text and images to produce an evocative and aesthetic description of the ethnographic research.
This iteration also served as verification of the correctness of the taxonomy.
%
The author's creation of and engagement with the presentation acted as a daily conversation with the research material. This allowed the author to concurrently and iteratively develop and articulate an understanding of the subject matter both visually and textually. At the end of the research period, the author engaged in a summative analysis~\citep{160940690400300403.pdf}  of the 
research material to review the completeness and consistency of the taxonomy.

\subsection{Self-Disclosure}%
While the author has experimented with text-to-image systems and produced digital artworks with these systems, the author is not an artist.
The author's background is in Computer Science with focus on Human-Computer Interaction (HCI) and Social Computing. The research was conducted not from a technical lens, but a human-centered lens~\citep{p32-guzdial.pdf}.
The author's specific interest in prompt engineering is the text-based interactions of users with text-to-image systems and the novel creative practices that arise from these systems.

\section{Taxonomy of Prompt Modifiers}%
\label{sec:modifiers}%
This research points towards there being 
six different types of prompt modifiers
    (subject terms,
    image prompts,
    style modifiers,
    quality boosters,
    repeating terms, and
    magic terms)
used by practitioners in the text-to-image art community (see summarized in \autoref{tab:modifiers}).
%
This taxonomy reflects the practitioner's comprehension of prompt modifiers,
a knowledge that was instrumental in classifying these modifiers into six distinct categories.

\begin{table}[htb]%
\caption{Taxonomy of prompt modifiers.}%
\label{tab:modifiers}%
\small
\begin{tabularx}{\textwidth}{lX}%
\toprule%
Modifier & Description  \\
\midrule%
Subject term      & Denotes the subject  \\
Style modifier    & Indicates an artistic style   \\
Image prompt      & Indicates a style or subject via an image   \\
Quality booster   & A term intended to improve the quality of the image   \\
Repeating term    & Repetition of subject terms or style terms with the intention of strengthening this subject or style    \\
Magic term        & A term that is semantically different from the rest of the prompt with the intention to produce surprising results   \\
\bottomrule%
\end{tabularx}%
\end{table}%

\textbf{Subject terms} indicate the desired subject to the text-to-image system (e.g., \textit{``a landscape''} or \textit{``an old car in a meadow''}).
While it is possible to generate images without subject terms, the subject is essential for controlling the image generation process.
On the other hand, since text-to-image systems were trained on images in context of their descriptive text, subject terms can, in some cases, have less control over the outcome.
	One such case is the artist Zdzisław Beksiński who developed a unique and recognizable style but never provided titles for his artworks. For this reason, early text-to-image systems, such as VQGAN--CLIP, struggled to reliably reproduce specific subjects in images generated to resemble Beksiński's artworks.

\textbf{Style modifiers} can be added to a prompt to produce images in a certain style.
Style modifiers will consistently reproduce a characteristic style (e.g., ``oil painting'') or artistic medium (e.g., ``mixed media'').
    For instance, the modifier \textit{``by Francisco Goya''} will generate digital images in the recognizable style of the late Spanish painter.
Other examples of this type of modifier include, but are not limited to,
    \textit{``oil on canvas,''}
    \textit{``\#pixelart,''}
    \textit{``hyperrealistic,''} 
    \textit{``abstract painting,''}
    \textit{``surreal,''}
    \textit{``Cubism''} or \textit{``cubist,''} 
    \textit{``cabinet card,''}
    \textit{``in the style of a cartoon,''}
    \textit{``by Claude Lorrain,''}
    and
    \textit{``in the style of Hudson River School,''} to name but a few.
As can be seen from the above list, style modifiers can include information about art periods, schools, and styles, but also art materials and media, techniques, and artists.
    When it comes to the latter, modifiers such as \textit{``by Greg Rutkowski''}  and \textit{``by James Gurney''} have become popular in the community of text-to-image art as a means 
    to produce images in a certain style and quality.

\textbf{Image prompts} act similar to subject terms and style modifiers in that they provide the text-to-image system a (visual) target for the synthesis of the image (both in terms of style and subject). Image prompts are typically specified as one or several urls that are added to the textual input prompt or provided in a separate array.
Image prompts are different from ``initial images'' which were investigated by \citet{3527927.3532792.pdf}.
Whereas an image prompt can consist of multiple images, there can only be one initial image. This initial image can be specified as a starting point for the image generation, for instance, for the purpose of enhancing or distorting the initial image. This is made possible because of the iterative nature of the image generation process which typically starts with an image filled with random noise (such as Perlin noise).

\textbf{Quality boosters} can be added to a prompt to increase aesthetic qualities and the level of detail in images.
Examples of this type of modifier are the terms 
    \textit{``trending on artstation,''}
    \textit{``award-winning,''}
    \textit{``masterpiece,''}
    \textit{``highly de\-tail\-ed}'',
    \textit{``awesome,''}
    \textit{``\#wow,''}
    \textit{``epic,''}
    and
    \textit{``ren\-der\-ed in Un\-real Engine.''}
%
This type of modifier can also be applied in the form of ``extra fluff'' added to the prompt.
Verbosity in the prompt may 
boost the amount of details and overall quality of the generated image, at the expense of the subject becoming less controllable.
    For instance, the prompt \textit{``painting of an exploding heart''} could potentially be improved by appending the modifiers 
    \textit{``highly detailed, eclectic, fiery, vfx, rendered in octane, postprocessing, 8k.''}

\textbf{Repeating terms} can 
strengthen the associations formed by the generative system.
For instance, the prompt \textit{``space whale. a whale in space''}\footnote{https://twitter.com/nshepperd1/status/1456584388037148678} by @nshepperd1
will likely produce subjectively better results than either of the two subject terms alone.
The use of different phrasing and synonyms will cause the text-to-image system to more reliably activate regions in the neural network's latent space that are associated with the subject terms. 
This is not only an imagined effect. The prompt \textit{``a very very very very very beautiful landscape''} will, for instance, likely produce a better image than a prompt without repeating terms. Technically, this is due to likelihood-maximizing language models becoming stuck in positive feedback loops from repeated phrases
\citep{1904.09751.pdf}. 

\textbf{Magic terms} introduce randomness 
to the image that can lead to surprising results. For instance, Twitter user
@jd\_pressman
added the magic term \textit{``control the soul''} to the prompt
\textit{``orchestra conductor leading a chorus of sound wave audio waveforms swirling around him on the orchestral stage''}.\footnote{https://twitter.com/jd\_pressman/status/1457171648293924867}
The term was added to~-- in Pressman's words~-- produce ``more magic, more wizard-ish imagery''.\footnote{https://twitter.com/jd\_pressman/status/1457445367125921793}
Magic terms, thus, introduce an element of unpredictability and surprise to the resulting images, often with the intention of increasing the variation in the output.
Magic terms can refer to terms that are semantically distant to the main subject of the prompt, or they can refer to non-visual qualities, such as the
    sense of touch (somatosensory),
    sense of hearing (auditory),
    sense of smell (olfactory),
    and
    sense of taste (gustatory)
(e.g., \textit{``feed the soul''} and \textit{``feel the sound''}).


In summary, prompt modifiers come in a variety of types and can take different forms. They can, for instance, be added as
	hash tags (e.g., \textit{``\#wow''}),
	attribution phrases (e.g., \textit{``by [artist]''}),
	or more complex composite statements (e.g., \textit{``in the style of [artist]''}).
Further, not every part of a prompt has the same importance and there are specific affordances of text-to-image systems that are being used in the practice of prompt engineering, as described in the following section.


\section{Prompt Engineering in Practice}%
\label{sec:promptengineering}%
%
This section provides an overview of how the different types of prompt modifiers are being applied in the practice of prompt engineering with specific focus on the generation of static images from either textual or visual input prompts.
We specifically focus on demonstrating  and explaining the iterative process of text-image generation with its iterative different steps (as described in \autoref{tab:modifiers}).

The first step in iterative prompt design is to denote the \textbf{subject} with one or several terms. While images can be generated from random text or even single characters and emojis~\citep{creativitypromptengineering}, the subject term is fundamental to the controlled generation of digital images.
Consequently, a prompt typically contains at least one subject term. Any other parts of the prompt are optional.
	It is, for instance, possible to generate artworks with the prompt \textit{``car.''} In practice, however, practitioners use modifiers to improve the resulting images and to exercise more control over the image creation process.

\textbf{Modifiers} are typically added with the intention to either modify the style of the generated image or boost its quality. As mentioned in Section~\ref{sec:modifiers}, style modifiers and quality boosters do not form a disparate set.
Rather, the two types of modifiers can have overlapping effects and the difference between the two types of prompt modifiers is sometimes not fully apparent.
For instance, the modifier \textit{``by Greg Rutkowski''} exhibits this 
property.
	Greg Rutkowski\footnote{https://www.artstation.com/rutkowski} is a  contemporary illustrator and concept artist who has been embraced by the text-to-image art community in their practice of prompt engineering.
	Images generated with the modifiers \textit{``by greg rutkowski''} or \textit{``in the style of greg rutkowski''} are of high quality, texture-rich, and contain a high amount of details.
As such, this modifier is often not used as a style modifier -- as one would expect --, but as a quality booster in the community, even though a trained eye may tell by the style of the image that the prompt modifier was being used.

Once a style modifier has been added, the style can be reinforced and ``solidified'' without losing expressivity.
\textbf{Solidifiers} (in the form of repeating terms)
can be applied to any of the other types of modifiers (subjects, style modifiers, and quality boosters), although they are most commonly applied to subject terms.
Image prompts are a special case in that they can carry both information about the subject and style because of their visual nature. If the textual prompt is aligned with the image prompt, the image prompt can also act as a solidifier.
On the other hand, if several images that are different from each other are added to the prompt, the image prompts will contribute to variation in the output.
Last, \textbf{magic terms} may be optionally added to increase the chance of surprising results.
The use of magic terms will result in more variation in the output, while maintaining the overall style.

Each of the six types of prompt modifiers can be assigned \textbf{weights}.
Weighted terms can be negative to exclude subjects and styles from being generated.
	For instance, VQGAN--CLIP tends to generate heart-shaped objects 
	when the prompt contains the word \textit{``love.''} By adding a negative weight to the prompt (e.g., \textit{``heart:-1''}), the system can be instructed not to activate the corresponding latents in its neural network. The resulting image is thus free from heart-shaped objects.
Weighted terms can also be used to seamlessly mix styles.
For instance, Twitter user @c0y0te6 mixed the styles of two artists in the prompt
	\textit{``a painting of a high prestess [sic] summoning a de\-mon by Ralph Mc\-Quar\-rie:75 $\vert$ by  Zdzislaw Bek\-sinski:25''}.\footnote{https://twitter.com/c0y0te6/status/1481780797858275329}
	The style of Ralph Mc\-Quar\-rie is, in this case, given precedence over the style of Zdzisław Beksiński (with a ratio of 3:1).

\autoref{tab:promptengineering} summarizes the iterative nature of prompt writing (c.f. \autoref{fig:promptengineering}).
Subject terms are most important for the controlled generation of images and usually written as first step. Modifiers and solidifers are then added to the prompt, either iteratively (image after image) or from learned experience.
Last, weights can be applied to exclude or mix subjects and styles.

\begin{table}[htb]%
\caption{The iterative practice of prompt writing.}%
\label{tab:promptengineering}%
\begin{tabularx}{\textwidth}{clXl}%
	\toprule%
	Step & Purpose & Prompt modifier & Importance \\
	\midrule%
	1 & Define & subject term, initial images, image prompt & required \\
	2 & Modify & style modifier, quality booster, initial images, image prompt & optional \\
	3 & Solidify & repeating terms, initial images  & optional \\
	4 & Vary  & magic terms, initial images  & optional \\
	5 & Mix/Exclude & mixing and exclusion  & optional \\
	\bottomrule%
\end{tabularx}%
\end{table}%

\section{Discussion}%
\label{sec:discussion}%
%
The availability and accessibility of text-to-image generation as a new creative practice and 
artistic medium \cite{2301.13049.pdf}, paired with a specific bundle of technologies and resources that support the ecosystem of this ``emerging art scene''~\citep{aliendreams,creativitypromptengineering}, have resulted in an explosion of AI generated artworks being shared online.
The application of prompt modifiers is key to the emerging creative practice called prompt engineering. 
The taxonomy of six different types of prompt modifiers represents an initial work to bringing structure to the creation process and research in text-to-image systems.
The taxonomy of prompt modifiers is reified for the sparse HCI literature around prompt engineering as a logical building block in this emerging field of research.

Midjourney has over 15 million members at the time of writing,\footnote{See https://discord.com/servers.}
and open source systems, such as 
Stable Diffusion, are available for execution on cloud or local hardware.
Today, everyone is able to synthesize digital images and artworks from natural language using free or relatively inexpensive means, with implications for productivity and creativity \cite{creativitypromptengineering}.
Gartner estimated in 2021 that by the year 2024, 80\% of technology products and services will be built by people who are not technology professionals~\citep{gartner}.
Increasingly, deep generative models will be used by laypeople without technical expertise and skills. Interaction with opaque deep learning models will increasingly become more common in future use cases and applications of artificial intelligence.
Therefore, prompt engineering is an emerging and important research area in the field of Human-Computer Interaction (HCI).
However, with the exception of the design guidelines by \citeauthor{chilton}~\citep{chilton} and \citeauthor{3527927.3532792.pdf}~\citep{3527927.3532792.pdf},
the scholarly literature in the field of HCI on prompt engineering still resembles a cottage industry, with concepts and structures yet to emerge.
Meanwhile, many resources started to emerge from within the online community, such as \citeauthor{travelersguide}'s ``Traveler's Guide to the Latent Space''~\citep{travelersguide} and \citeauthor{dallepromptbook}'s ``DALL-E Prompt Book''~\citep{dallepromptbook}.
Drawing on gray literature, such as the above, and extensive auto-ethnographic research, this work provides a taxonomy of prompt modifiers as a starting point for systematizing the practice of prompt engineering for text-to-image generation.
The subsequent discussion will examine the broader implications of prompt engineering for human-AI interaction.

\subsection{Broader Implications for Human-AI Interaction}%
Research on prompt engineering has broader implications
and is not only limited to the field of text-to-image synthesis and AI generated art, but also relevant to the interaction of humans with deep learning models and artificial intelligence in general.


\subsubsection{AI and the future of creative work}%
There is much potential for deep learning to disrupt and transform entire sectors of the creative economy.
Recently, there has been an interest into developing generative systems that are able to synthesize more complex outcomes. For instance, systems for text-to-video generation have been presented by \citet{2205.15868.pdf}, \citet{imagen-video.pdf}, \citet{2209.14792.pdf}, and \citet{phenaki.pdf}.
Low-code and no-code tools for creating online products and experiences will become increasingly common in the future. Declarative machine learning systems may~-- as a next wave of machine learning~-- bring machine learning to non-coders~\citep{3475167.pdf}.
This technology will extend the currently rather narrow focus of prompt engineering on language models and text-to-image synthesis to more broader application domains. 
In the future, we may see deep generative models with generative capabilities that transcend what we can imagine today.
    Deep generative models could, for instance, create entire interactive story-driven worlds and games from short text prompts.

Such powerful AI-based systems will have implications for the future of creative work.
Artificial intelligence will not only transform the way we interact with computers and perform work online, but also the content of our work and the human agency in the work.
An example of an application that has such transformative potential is OpenAI's Codex~\citep{chen2021evaluating,codex}.
    Codex is a large language model that interprets commands in natural language and generates programming code.
        In the future, instead of typing code, we will be able to describe a software and its expected outputs in natural language.
        Pre-trained generative models, such as
        Codex, BLOOM \citep{BLOOM}, or other ``foundation models''~\citep{foundationmodels}, will then generate executable software code based on the human's spoken or written input prompts.
    This technology has already found application in GitHub's CoPilot\footnote{copilot.github.com}, an ``AI pair programmer'' assisting its users in auto-completing programming code.
    In academia, researchers increasingly rely on language models as creativity support tools for writing academic papers \citep{d41586-022-03479-w.pdf}.
    The change in the agency of humans and computers brought by generative models will be transformative to creative work, such as software development and research.


\subsubsection{Beyond text-to-image generation}%
The use case of art generated with text-to-image systems discussed in this paper is but one of many application areas of prompt engineering, with implications for the future of creative work and Human-AI Interaction (HAI) in general.
    The latter can be viewed from many different perspectives, such as human-centered AI~\citep{2002.04087.pdf}, human-AI partnerships~\citep{RAMCHURN2021102891}, and human-AI cooperation~\citep{s41467-017-02597-8.pdf}.
Irrespective of the term used to describe our relationship with AI, we will increasingly interact with opaque models through prompts in natural language.

Research on how to design prompts is therefore timely and important.
Increasingly, we see use-facing applications being powered by foundation-scale models. The emergent properties of these models make it possible to use them for a vast number of different use cases and applications.
Internally, such applications are often enabled by prompt engineering.
For instance, tool-augmented language models \cite{2302.07842.pdf,AutoGPT} internally use prompts to enable the language model to use external tools.
Research on prompt engineering, thus, will advance our understanding of how people can effectively interact with and employ machine learning models for solving complex tasks.

With these considerations in mind, we turn our attention to specific opportunities and challenges of prompt engineering within the field of Human-Computer Interaction (HCI).

\subsection{Opportunities for Research on Prompt Engineering in HCI}%


This section discusses opportunities for future research on prompt engineering in the field of HCI.
Specifically,
we touch on the community dynamics surrounding text-to-image art creation,
novel workflows and techniques employed by practitioners, and
the embedded biases in AI-driven systems.
Additionally, we explore the relevance of prompt engineering for research on computational aesthetics and human-AI alignment.

\subsubsection{Social aspects of prompt engineering}
There are social components to the use of text-to-image generation systems.
Prompt engineers face an interesting challenge:
Because text-to-image systems were trained on images and text
scraped from the Web, users of text-to-image systems need to imagine and predict how other people described and reacted to images posted on the Web.
Describing an image in detail is often not enough to achieve optimal results~-- one has to imagine the image as if it already existed on the Web.

Another social aspect in prompt engineering are the dedicated communities that came into existence only recently.
Practitioners of text-to-image art are producing artworks in shared Discord-based chat rooms, such as on Midjourney.\footnote{https://www.midjourney.com}
These dedicated communities offer a rich set of social features worth investigating more closely in HCI research.
For instance, members on Midjourney have their own profiles that bundle the members' successful creations together with the prompts used to create the images. Midjourney introduced a 2D map in which members can explore other members based on the similarity of their prompts.
Midjourney also has dedicated ``group jam'' sections in which members can iterate on and further develop other members' works and there is a ``theme of the day'' section. Long running threads are quite common in this community.
Community-learning is an interesting area of research in this regard.
How do members receive and seek inspiration in the community? How do novices learn the craft of prompt engineering and is there learning taking place in the community as a whole?

Future work could explore and ethnographically investigate the online community around text-to-image art and its prompt engineering practices in more detail, using the taxonomy presented in this paper as a conceptual starting point or framework.

\subsubsection{Human-AI co-creation}
While the heart piece of prompt engineering is prompt writing, prompt engineering is only a starting point in some practitioners' creative work flows.
Novel creative practices are emerging.
    For instance, practitioners may develop complex work flows for creating their artworks (e.g., generating initial images with one text-to-image system as a source for inspiration, then continuing on another text-to-image system before finalizing the images in a photo editor).
    The different affordances of text-to-image systems still need to be reified and systematized in the HCI community.
    For instance, some text-to-image systems enable the creation of zooming animations, others can complete parts of images which is called image inpainting~\citep{3394171.3414017.pdf} and outpainting.\footnote{See, for instance, https://twitter.com/adampickard/status/1551584412659335168.}
    These novel creative practices offer a level of interactivity beyond mere generation of static images from textual input prompts.
    Further, practitioners may make certain idiosyncratic choices when they create text-based generative art (e.g., selecting certain numerical values as seed for the model or adapting the canvas size to certain subject terms).
    Some of these choices may fall into the realm of 
    folk theories \citep{2858036.2858494.pdf,nihms-441164.pdf}~-- that is, causal attributions that may or may not be true~--, while other choices may be based on the practitioner's experimentation and experience with prompt engineering.
Future work could investigate these creative practices, work flows, strategies, and beliefs adopted by practitioners in the text-to-image art community.
The emerging research field also offers an opportunity for HCI researchers to make technical contributions~\citep{wobbrock} in the form of creativity support tools, user interfaces, and interactive experiences to support text-to-image generation, to teach novices the practice of prompt engineering, and to advance the emerging AI generated art ecosystem.
Research in this space could make a timely contribution to a novel computational medium and an emerging digital art form.

\subsubsection{Bias in image generation systems}
Another interesting area for future work is bias encoded in text-to-image generation systems.
It has been shown, for instance, that
the CLIP model contains bias\footnote{See https://twitter.com/RiversHaveWings/status/1432100170645180416.} and 
some text-to-image systems prompted with \textit{``princess''} will produce images of women with light skin color, reflecting the bias in the training data toward Western, educated, industrialized, rich and democratic (WEIRD) subjects.\footnote{See https://twitter.com/EMostaque/status/1495323912951021568.}
OpenAI recently announced that bias was reduced in their DALL-E~2 model~\citep{dalle2bias}, but at the cost of potentially reducing signal-to-noise of the generated images.\footnote{See https://twitter.com/minimaxir/status/1549070583035416576.}

Responsible deployment of large models
and the potential risks are two concerns often listed for not fully releasing a model.
While organizations such as OpenAI and Google can be commended for trying to be responsible with their powerful systems, these organizations act paternalistic and impose their value and belief system onto their users which is another source of bias.
DALL-E~2, in particular, can be a source of frustration for its users who are often faced with content policy notices for terms relating to war or sexual content (with a  threat of account closure if the warning is incurred too often).
\citeauthor{pressmancrowson2022} recently raised an important point:
Humans are sexual beings and the androgynous values imposed on text-to-image systems with the intent of making them ``safe-for-work'' deprives users of ``a key component of human aesthetic values and experience''~\citep{pressmancrowson2022}.

\subsubsection{Computational aesthetics and Human-AI alignment}%
The goal of making computers evaluate and understand aesthetics is much older than text-to-image generation~\citep{Galanter2012_Chapter_ComputationalAestheticEvaluati.pdf}.
Recently, there is renewed research on neural image assessment and computational aesthetics.
State-of-the-art text-to-image systems increasingly consider human aesthetics in an attempt to produce better images \citep{laion-aesthetics}.
Prompts are a vast resource for research on computational aesthetics, as they encapsulate a person's stated intent. This intent, however, is likely only partially explicit.
Research on prompt engineering, therefore, also relates to research on human-AI alignment \citep{Gabriel2020_Article_ArtificialIntelligenceValuesAn.pdf}.
This research area is concerned with teaching artificial intelligence to understand human values.
Prompts for text-to-image generation systems could form an interesting study resource for this kind of research.

\section{Conclusion}%
\label{sec:conclusion}%


This research contributes to the academic understanding of text-to-image generation by proposing a novel taxonomy of six types of prompt modifiers:
    subject terms,
    image prompts,
    style modifiers,
    quality boosters,
    repeating terms, and
    magic terms.
The taxonomy 
of prompt modifiers lays the foundation for future structured investigations into prompt engineering for text-to-image generation and AI generated art.
Moreover, the taxonomy highlights the unique affordances of text-to-image systems, providing a clearer understanding of the design (``engineering'') of prompts for image generation.

In the practice of prompt engineering for generating static images from textual or visual inputs, subject terms are fundamental to the controlled creation of images. Practitioners often use prompt modifiers to improve image quality and exercise greater control over the creation process. Modifiers either modify the image style or enhance its quality, and these two types can overlap in their effects. For example, the modifier ``by Greg Rutkowski'' is typically used by practitioners as a quality booster rather than a style modifier, despite the artist's distinct style. Solidifiers can also be used to reinforce a chosen style or subject without loss of expressivity. Image prompts, due to their visual nature, can carry information about both subject and style. The use of magic terms can increase output variation while maintaining style. Additionally, prompt modifiers can be assigned weights to control image generation further. Negative weights can exclude certain subjects or styles, while positive weights can be used to mix styles. The process of prompt writing is iterative, starting with subject terms, followed by the addition of modifiers and solidifiers, and finally applying weights for precise control.

This work has illuminated the burgeoning field of prompt engineering, which is central to the emerging practice of text-to-image synthesis and AI-generated art. The development of a taxonomy of six types of prompt modifiers is a stepping stone to bring structure to this area of study.
Future research will be critical in addressing several key areas, including the ethical and societal implications of AI-generated creative work, the social aspects of prompt engineering, the co-creation process between humans and AI, potential bias in image generation systems, and the alignment of AI with human values. As non-technical users increasingly interact with complex AI models, the need for HCI research in prompt engineering will only continue to grow. The exploration of these topics will not only advance our understanding of how people can effectively interact with machine learning models, but also inform the design of future AI-driven systems and contribute to the development of a novel digital art form.

\clearpage

\bibliographystyle{ACM-Reference-Format}%
\bibliography{paper}%


\begin{thebibliography}{60}


\ifx \showCODEN    \undefined \def \showCODEN     #1{\unskip}     \fi
\ifx \showDOI      \undefined \def \showDOI       #1{#1}\fi
\ifx \showISBNx    \undefined \def \showISBNx     #1{\unskip}     \fi
\ifx \showISBNxiii \undefined \def \showISBNxiii  #1{\unskip}     \fi
\ifx \showISSN     \undefined \def \showISSN      #1{\unskip}     \fi
\ifx \showLCCN     \undefined \def \showLCCN      #1{\unskip}     \fi
\ifx \shownote     \undefined \def \shownote      #1{#1}          \fi
\ifx \showarticletitle \undefined \def \showarticletitle #1{#1}   \fi
\ifx \showURL      \undefined \def \showURL       {\relax}        \fi
\providecommand\bibfield[2]{#2}
\providecommand\bibinfo[2]{#2}
\providecommand\natexlab[1]{#1}
\providecommand\showeprint[2][]{arXiv:#2}

\bibitem[{ArtHub}(2022)]%
        {arthub.ai}
\bibfield{author}{\bibinfo{person}{{ArtHub}}.} \bibinfo{year}{2022}\natexlab{}.
\newblock \bibinfo{title}{arthub.ai}.
\newblock
\newblock
\newblock
\shownote{{https}://arthub.ai/ [Accessed Nov. 9, 2022]}.


\bibitem[Bach et~al\mbox{.}(2022)]%
        {2022.acl-demo.9.pdf}
\bibfield{author}{\bibinfo{person}{Stephen Bach}, \bibinfo{person}{Victor
  Sanh}, \bibinfo{person}{Zheng~Xin Yong}, \bibinfo{person}{Albert Webson},
  \bibinfo{person}{Colin Raffel}, \bibinfo{person}{Nihal~V. Nayak},
  \bibinfo{person}{Abheesht Sharma}, \bibinfo{person}{Taewoon Kim},
  \bibinfo{person}{M~Saiful Bari}, \bibinfo{person}{Thibault Fevry},
  \bibinfo{person}{Zaid Alyafeai}, \bibinfo{person}{Manan Dey},
  \bibinfo{person}{Andrea Santilli}, \bibinfo{person}{Zhiqing Sun},
  \bibinfo{person}{Srulik Ben-david}, \bibinfo{person}{Canwen Xu},
  \bibinfo{person}{Gunjan Chhablani}, \bibinfo{person}{Han Wang},
  \bibinfo{person}{Jason Fries}, \bibinfo{person}{Maged Al-shaibani},
  \bibinfo{person}{Shanya Sharma}, \bibinfo{person}{Urmish Thakker},
  \bibinfo{person}{Khalid Almubarak}, \bibinfo{person}{Xiangru Tang},
  \bibinfo{person}{Dragomir Radev}, \bibinfo{person}{Mike Tian-jian Jiang},
  {and} \bibinfo{person}{Alexander Rush}.} \bibinfo{year}{2022}\natexlab{}.
\newblock \showarticletitle{{P}rompt{S}ource: An Integrated Development
  Environment and Repository for Natural Language Prompts}. In
  \bibinfo{booktitle}{\emph{Proceedings of the 60th Annual Meeting of the
  Association for Computational Linguistics: System Demonstrations}}.
  \bibinfo{publisher}{Association for Computational Linguistics},
  \bibinfo{address}{Dublin, Ireland}, \bibinfo{pages}{93--104}.
\newblock
\urldef\tempurl%
\url{https://doi.org/10.18653/v1/2022.acl-demo.9}
\showDOI{\tempurl}


\bibitem[{BigScience Initiative}(2022)]%
        {BLOOM}
\bibfield{author}{\bibinfo{person}{{BigScience Initiative}}.}
  \bibinfo{year}{2022}\natexlab{}.
\newblock \showarticletitle{Introducing The World's Largest Open Multilingual
  Language Model: {BLOOM}.}
\newblock  (\bibinfo{year}{2022}).
\newblock
\newblock
\shownote{{https}://bigscience.huggingface.co/blog/bloom [Accessed Nov. 9,
  2022].}.


\bibitem[Bommasani et~al\mbox{.}(2021)]%
        {foundationmodels}
\bibfield{author}{\bibinfo{person}{Rishi Bommasani}, \bibinfo{person}{Drew~A.
  Hudson}, \bibinfo{person}{Ehsan Adeli}, \bibinfo{person}{Russ Altman},
  \bibinfo{person}{Simran Arora}, \bibinfo{person}{Sydney von Arx},
  \bibinfo{person}{Michael~S. Bernstein}, \bibinfo{person}{Jeannette Bohg},
  \bibinfo{person}{Antoine Bosselut}, \bibinfo{person}{Emma Brunskill},
  \bibinfo{person}{Erik Brynjolfsson}, \bibinfo{person}{Shyamal Buch},
  \bibinfo{person}{Dallas Card}, \bibinfo{person}{Rodrigo Castellon},
  \bibinfo{person}{Niladri Chatterji}, \bibinfo{person}{Annie Chen},
  \bibinfo{person}{Kathleen Creel}, \bibinfo{person}{Jared~Quincy Davis},
  \bibinfo{person}{Dora Demszky}, \bibinfo{person}{Chris Donahue},
  \bibinfo{person}{Moussa Doumbouya}, \bibinfo{person}{Esin Durmus},
  \bibinfo{person}{Stefano Ermon}, \bibinfo{person}{John Etchemendy},
  \bibinfo{person}{Kawin Ethayarajh}, \bibinfo{person}{Li Fei-Fei},
  \bibinfo{person}{Chelsea Finn}, \bibinfo{person}{Trevor Gale},
  \bibinfo{person}{Lauren Gillespie}, \bibinfo{person}{Karan Goel},
  \bibinfo{person}{Noah Goodman}, \bibinfo{person}{Shelby Grossman},
  \bibinfo{person}{Neel Guha}, \bibinfo{person}{Tatsunori Hashimoto},
  \bibinfo{person}{Peter Henderson}, \bibinfo{person}{John Hewitt},
  \bibinfo{person}{Daniel~E. Ho}, \bibinfo{person}{Jenny Hong},
  \bibinfo{person}{Kyle Hsu}, \bibinfo{person}{Jing Huang},
  \bibinfo{person}{Thomas Icard}, \bibinfo{person}{Saahil Jain},
  \bibinfo{person}{Dan Jurafsky}, \bibinfo{person}{Pratyusha Kalluri},
  \bibinfo{person}{Siddharth Karamcheti}, \bibinfo{person}{Geoff Keeling},
  \bibinfo{person}{Fereshte Khani}, \bibinfo{person}{Omar Khattab},
  \bibinfo{person}{Pang~Wei Koh}, \bibinfo{person}{Mark Krass},
  \bibinfo{person}{Ranjay Krishna}, \bibinfo{person}{Rohith Kuditipudi},
  \bibinfo{person}{Ananya Kumar}, \bibinfo{person}{Faisal Ladhak},
  \bibinfo{person}{Mina Lee}, \bibinfo{person}{Tony Lee}, \bibinfo{person}{Jure
  Leskovec}, \bibinfo{person}{Isabelle Levent}, \bibinfo{person}{Xiang~Lisa
  Li}, \bibinfo{person}{Xuechen Li}, \bibinfo{person}{Tengyu Ma},
  \bibinfo{person}{Ali Malik}, \bibinfo{person}{Christopher~D. Manning},
  \bibinfo{person}{Suvir Mirchandani}, \bibinfo{person}{Eric Mitchell},
  \bibinfo{person}{Zanele Munyikwa}, \bibinfo{person}{Suraj Nair},
  \bibinfo{person}{Avanika Narayan}, \bibinfo{person}{Deepak Narayanan},
  \bibinfo{person}{Ben Newman}, \bibinfo{person}{Allen Nie},
  \bibinfo{person}{Juan~Carlos Niebles}, \bibinfo{person}{Hamed Nilforoshan},
  \bibinfo{person}{Julian Nyarko}, \bibinfo{person}{Giray Ogut},
  \bibinfo{person}{Laurel Orr}, \bibinfo{person}{Isabel Papadimitriou},
  \bibinfo{person}{Joon~Sung Park}, \bibinfo{person}{Chris Piech},
  \bibinfo{person}{Eva Portelance}, \bibinfo{person}{Christopher Potts},
  \bibinfo{person}{Aditi Raghunathan}, \bibinfo{person}{Rob Reich},
  \bibinfo{person}{Hongyu Ren}, \bibinfo{person}{Frieda Rong},
  \bibinfo{person}{Yusuf Roohani}, \bibinfo{person}{Camilo Ruiz},
  \bibinfo{person}{Jack Ryan}, \bibinfo{person}{Christopher Ré},
  \bibinfo{person}{Dorsa Sadigh}, \bibinfo{person}{Shiori Sagawa},
  \bibinfo{person}{Keshav Santhanam}, \bibinfo{person}{Andy Shih},
  \bibinfo{person}{Krishnan Srinivasan}, \bibinfo{person}{Alex Tamkin},
  \bibinfo{person}{Rohan Taori}, \bibinfo{person}{Armin~W. Thomas},
  \bibinfo{person}{Florian Tramèr}, \bibinfo{person}{Rose~E. Wang},
  \bibinfo{person}{William Wang}, \bibinfo{person}{Bohan Wu},
  \bibinfo{person}{Jiajun Wu}, \bibinfo{person}{Yuhuai Wu},
  \bibinfo{person}{Sang~Michael Xie}, \bibinfo{person}{Michihiro Yasunaga},
  \bibinfo{person}{Jiaxuan You}, \bibinfo{person}{Matei Zaharia},
  \bibinfo{person}{Michael Zhang}, \bibinfo{person}{Tianyi Zhang},
  \bibinfo{person}{Xikun Zhang}, \bibinfo{person}{Yuhui Zhang},
  \bibinfo{person}{Lucia Zheng}, \bibinfo{person}{Kaitlyn Zhou}, {and}
  \bibinfo{person}{Percy Liang}.} \bibinfo{year}{2021}\natexlab{}.
\newblock \bibinfo{booktitle}{\emph{On the Opportunities and Risks of
  Foundation Models}}.
\newblock \bibinfo{type}{{T}echnical {R}eport}. \bibinfo{institution}{Stanford
  University}.
\newblock
\urldef\tempurl%
\url{https://crfm.stanford.edu/assets/report.pdf}
\showURL{%
\tempurl}


\bibitem[Chen et~al\mbox{.}(2021)]%
        {chen2021evaluating}
\bibfield{author}{\bibinfo{person}{Mark Chen}, \bibinfo{person}{Jerry Tworek},
  \bibinfo{person}{Heewoo Jun}, \bibinfo{person}{Qiming Yuan},
  \bibinfo{person}{Henrique~Ponde de Oliveira~Pinto}, \bibinfo{person}{Jared
  Kaplan}, \bibinfo{person}{Harri Edwards}, \bibinfo{person}{Yuri Burda},
  \bibinfo{person}{Nicholas Joseph}, \bibinfo{person}{Greg Brockman},
  \bibinfo{person}{Alex Ray}, \bibinfo{person}{Raul Puri},
  \bibinfo{person}{Gretchen Krueger}, \bibinfo{person}{Michael Petrov},
  \bibinfo{person}{Heidy Khlaaf}, \bibinfo{person}{Girish Sastry},
  \bibinfo{person}{Pamela Mishkin}, \bibinfo{person}{Brooke Chan},
  \bibinfo{person}{Scott Gray}, \bibinfo{person}{Nick Ryder},
  \bibinfo{person}{Mikhail Pavlov}, \bibinfo{person}{Alethea Power},
  \bibinfo{person}{Lukasz Kaiser}, \bibinfo{person}{Mohammad Bavarian},
  \bibinfo{person}{Clemens Winter}, \bibinfo{person}{Philippe Tillet},
  \bibinfo{person}{Felipe~Petroski Such}, \bibinfo{person}{Dave Cummings},
  \bibinfo{person}{Matthias Plappert}, \bibinfo{person}{Fotios Chantzis},
  \bibinfo{person}{Elizabeth Barnes}, \bibinfo{person}{Ariel Herbert-Voss},
  \bibinfo{person}{William~Hebgen Guss}, \bibinfo{person}{Alex Nichol},
  \bibinfo{person}{Alex Paino}, \bibinfo{person}{Nikolas Tezak},
  \bibinfo{person}{Jie Tang}, \bibinfo{person}{Igor Babuschkin},
  \bibinfo{person}{Suchir Balaji}, \bibinfo{person}{Shantanu Jain},
  \bibinfo{person}{William Saunders}, \bibinfo{person}{Christopher Hesse},
  \bibinfo{person}{Andrew~N. Carr}, \bibinfo{person}{Jan Leike},
  \bibinfo{person}{Josh Achiam}, \bibinfo{person}{Vedant Misra},
  \bibinfo{person}{Evan Morikawa}, \bibinfo{person}{Alec Radford},
  \bibinfo{person}{Matthew Knight}, \bibinfo{person}{Miles Brundage},
  \bibinfo{person}{Mira Murati}, \bibinfo{person}{Katie Mayer},
  \bibinfo{person}{Peter Welinder}, \bibinfo{person}{Bob McGrew},
  \bibinfo{person}{Dario Amodei}, \bibinfo{person}{Sam McCandlish},
  \bibinfo{person}{Ilya Sutskever}, {and} \bibinfo{person}{Wojciech Zaremba}.}
  \bibinfo{year}{2021}\natexlab{}.
\newblock \showarticletitle{Evaluating Large Language Models Trained on Code.}
\newblock  (\bibinfo{year}{2021}).
\newblock
\showeprint[arxiv]{2107.03374}~[cs.LG]
\newblock
\shownote{[Preprint]. Available at: https://arxiv.org/abs/2107.03374 [Accessed
  Nov. 9, 2022].}.


\bibitem[Crandall et~al\mbox{.}(2018)]%
        {s41467-017-02597-8.pdf}
\bibfield{author}{\bibinfo{person}{Jacob~W. Crandall}, \bibinfo{person}{Mayada
  Oudah}, \bibinfo{person}{Tennom}, \bibinfo{person}{Fatimah Ishowo-Oloko},
  \bibinfo{person}{Sherief Abdallah}, \bibinfo{person}{Jean-François
  Bonnefon}, \bibinfo{person}{Manuel Cebrian}, \bibinfo{person}{Azim Shariff},
  \bibinfo{person}{Michael~A. Goodrich}, {and} \bibinfo{person}{Iyad Rahwan}.}
  \bibinfo{year}{2018}\natexlab{}.
\newblock \showarticletitle{Cooperating with machines}.
\newblock \bibinfo{journal}{\emph{Nature Communications}} \bibinfo{volume}{9},
  \bibinfo{number}{1} (\bibinfo{year}{2018}), \bibinfo{numpages}{12}~pages.
\newblock
\urldef\tempurl%
\url{https://doi.org/10.1038/s41467-017-02597-8}
\showDOI{\tempurl}


\bibitem[Crawford(1996)]%
        {crawford1996.pdf}
\bibfield{author}{\bibinfo{person}{Lyall Crawford}.}
  \bibinfo{year}{1996}\natexlab{}.
\newblock \showarticletitle{Personal ethnography}.
\newblock \bibinfo{journal}{\emph{Communication Monographs}}
  \bibinfo{volume}{63}, \bibinfo{number}{2} (\bibinfo{year}{1996}),
  \bibinfo{pages}{158--170}.
\newblock
\urldef\tempurl%
\url{https://doi.org/10.1080/03637759609376384}
\showDOI{\tempurl}


\bibitem[Crowson(2021)]%
        {CLIPguideddiffusion}
\bibfield{author}{\bibinfo{person}{Katherine Crowson}.}
  \bibinfo{year}{2021}\natexlab{}.
\newblock \showarticletitle{CLIP Guided Diffusion {HQ} 256x256}.
\newblock  (\bibinfo{year}{2021}).
\newblock
\newblock
\shownote{{https}://colab.research.google.com/drive/12a\_Wrfi2\_gwwAuN3VvMTwVMz9TfqctNj
  [Accessed Nov. 9, 2022]}.


\bibitem[Crowson et~al\mbox{.}(2022)]%
        {VQGANCLIP}
\bibfield{author}{\bibinfo{person}{Katherine Crowson}, \bibinfo{person}{Stella
  Biderman}, \bibinfo{person}{Daniel Kornis}, \bibinfo{person}{Dashiell
  Stander}, \bibinfo{person}{Eric Hallahan}, \bibinfo{person}{Louis
  Castricato}, {and} \bibinfo{person}{Edward Raff}.}
  \bibinfo{year}{2022}\natexlab{}.
\newblock \showarticletitle{VQGAN-CLIP: Open Domain Image Generation
  and Editing with Natural Language Guidance}. In
  \bibinfo{booktitle}{\emph{Computer Vision -- ECCV 2022}},
  \bibfield{editor}{\bibinfo{person}{Shai Avidan}, \bibinfo{person}{Gabriel
  Brostow}, \bibinfo{person}{Moustapha Ciss{\'e}},
  \bibinfo{person}{Giovanni~Maria Farinella}, {and} \bibinfo{person}{Tal
  Hassner}} (Eds.). \bibinfo{publisher}{Springer Nature},
  \bibinfo{address}{Cham, Switzerland}, \bibinfo{pages}{88--105}.
\newblock
\showISBNx{978-3-031-19836-6}


\bibitem[Denzin and Lincoln(2017)]%
        {handbook}
\bibfield{author}{\bibinfo{person}{Norman~K. Denzin} {and}
  \bibinfo{person}{Yvonna~S. Lincoln}.} \bibinfo{year}{2017}\natexlab{}.
\newblock \bibinfo{booktitle}{\emph{The SAGE Handbook of Qualitative Research}
  (\bibinfo{edition}{5th} ed.)}.
\newblock \bibinfo{publisher}{SAGE}, \bibinfo{address}{Thousand Oaks, CA}.
\newblock


\bibitem[Duncan(2004)]%
        {160940690400300403.pdf}
\bibfield{author}{\bibinfo{person}{Margot Duncan}.}
  \bibinfo{year}{2004}\natexlab{}.
\newblock \showarticletitle{Autoethnography: Critical Appreciation of an
  Emerging Art}.
\newblock \bibinfo{journal}{\emph{International Journal of Qualitative
  Methods}} \bibinfo{volume}{3}, \bibinfo{number}{4} (\bibinfo{year}{2004}),
  \bibinfo{pages}{28--39}.
\newblock
\urldef\tempurl%
\url{https://doi.org/10.1177/160940690400300403}
\showDOI{\tempurl}


\bibitem[Durant(2021)]%
        {artstudies}
\bibfield{author}{\bibinfo{person}{Remi Durant}.}
  \bibinfo{year}{2021}\natexlab{}.
\newblock \showarticletitle{Artist Studies by @remi\_durant}.
\newblock  (\bibinfo{year}{2021}).
\newblock
\newblock
\shownote{{https}://remidurant.com/artists/ [Accessed Nov. 9, 2022]}.


\bibitem[Ellis et~al\mbox{.}(2011)]%
        {10.2307_23032294.pdf}
\bibfield{author}{\bibinfo{person}{Carolyn Ellis}, \bibinfo{person}{Tony~E.
  Adams}, {and} \bibinfo{person}{Arthur~P. Bochner}.}
  \bibinfo{year}{2011}\natexlab{}.
\newblock \showarticletitle{Autoethnography: An Overview}.
\newblock \bibinfo{journal}{\emph{Historical Social Research / Historische
  Sozialforschung}} \bibinfo{volume}{36}, \bibinfo{number}{4 (138)}
  (\bibinfo{year}{2011}), \bibinfo{pages}{273--290}.
\newblock
\showISSN{01726404}
\urldef\tempurl%
\url{http://www.jstor.org/stable/23032294}
\showURL{%
\tempurl}


\bibitem[Eslami et~al\mbox{.}(2016)]%
        {2858036.2858494.pdf}
\bibfield{author}{\bibinfo{person}{Motahhare Eslami}, \bibinfo{person}{Karrie
  Karahalios}, \bibinfo{person}{Christian Sandvig}, \bibinfo{person}{Kristen
  Vaccaro}, \bibinfo{person}{Aimee Rickman}, \bibinfo{person}{Kevin Hamilton},
  {and} \bibinfo{person}{Alex Kirlik}.} \bibinfo{year}{2016}\natexlab{}.
\newblock \showarticletitle{First I "like" It, Then I Hide It: Folk Theories of
  Social Feeds}. In \bibinfo{booktitle}{\emph{Proceedings of the 2016 CHI
  Conference on Human Factors in Computing Systems}}
  \emph{(\bibinfo{series}{CHI '16})}. \bibinfo{publisher}{Association for
  Computing Machinery}, \bibinfo{address}{New York, NY},
  \bibinfo{pages}{2371–2382}.
\newblock
\showISBNx{9781450333627}
\urldef\tempurl%
\url{https://doi.org/10.1145/2858036.2858494}
\showDOI{\tempurl}


\bibitem[Gabha(2022)]%
        {DiscoDiffusionArtiststudies}
\bibfield{author}{\bibinfo{person}{Harmeet Gabha}.}
  \bibinfo{year}{2022}\natexlab{}.
\newblock \showarticletitle{Disco Diffusion 70+ Artist Studies}.
\newblock  (\bibinfo{year}{2022}).
\newblock
\newblock
\shownote{{https}://weirdwonderfulai.art/resources/disco-diffusion-70-plus-artist-studies/
  [Accessed Nov. 9, 2022].}.


\bibitem[Gabriel(2020)]%
        {Gabriel2020_Article_ArtificialIntelligenceValuesAn.pdf}
\bibfield{author}{\bibinfo{person}{Iason Gabriel}.}
  \bibinfo{year}{2020}\natexlab{}.
\newblock \showarticletitle{Artificial Intelligence, Values, and Alignment}.
\newblock \bibinfo{journal}{\emph{Minds and Machines}} \bibinfo{volume}{30},
  \bibinfo{number}{3} (\bibinfo{year}{2020}), \bibinfo{pages}{411--437}.
\newblock
\urldef\tempurl%
\url{https://doi.org/10.1007/s11023-020-09539-2}
\showDOI{\tempurl}


\bibitem[Galanter(2012)]%
        {Galanter2012_Chapter_ComputationalAestheticEvaluati.pdf}
\bibfield{author}{\bibinfo{person}{Philip Galanter}.}
  \bibinfo{year}{2012}\natexlab{}.
\newblock \bibinfo{booktitle}{\emph{Computational Aesthetic Evaluation: Past
  and Future}}.
\newblock \bibinfo{publisher}{Springer Berlin Heidelberg},
  \bibinfo{address}{Berlin, Heidelberg}, \bibinfo{pages}{255--293}.
\newblock
\showISBNx{978-3-642-31727-9}
\urldef\tempurl%
\url{https://doi.org/10.1007/978-3-642-31727-9_10}
\showDOI{\tempurl}


\bibitem[{Gartner}(2021)]%
        {gartner}
\bibfield{author}{\bibinfo{person}{{Gartner}}.}
  \bibinfo{year}{2021}\natexlab{}.
\newblock \showarticletitle{Gartner Says the Majority of Technology Products
  and Services Will Be Built by Professionals Outside of IT by 2024.}
\newblock \bibinfo{howpublished}{Press release}.
\newblock  (\bibinfo{date}{14 June} \bibinfo{year}{2021}).
\newblock
\newblock
\shownote{{https}://www.gartner.com/en/newsroom/press-releases/2021-06-10-gartner-says-the-majority-of-technology-products-and-services-will-be-built-by-professionals-outside-of-it-by-2024
  [Accessed Nov. 9, 2022]}.


\bibitem[Gelman and Legare(2011)]%
        {nihms-441164.pdf}
\bibfield{author}{\bibinfo{person}{Susan~A. Gelman} {and}
  \bibinfo{person}{Cristine~H. Legare}.} \bibinfo{year}{2011}\natexlab{}.
\newblock \showarticletitle{Concepts and folk theories}.
\newblock \bibinfo{journal}{\emph{Annual Review of Anthropology}}
  \bibinfo{volume}{40} (\bibinfo{year}{2011}), \bibinfo{pages}{379--398}.
\newblock
\urldef\tempurl%
\url{https://doi.org/10.1146/annurev-anthro-081309-145822}
\showDOI{\tempurl}


\bibitem[Gold(1958)]%
        {2573808.pdf}
\bibfield{author}{\bibinfo{person}{Raymond~L. Gold}.}
  \bibinfo{year}{1958}\natexlab{}.
\newblock \showarticletitle{Roles in Sociological Field Observations}.
\newblock \bibinfo{journal}{\emph{Social Forces}} \bibinfo{volume}{36},
  \bibinfo{number}{3} (\bibinfo{year}{1958}), \bibinfo{pages}{217--223}.
\newblock
\urldef\tempurl%
\url{http://www.jstor.org/stable/2573808}
\showURL{%
\tempurl}


\bibitem[Guzdial(2013)]%
        {p32-guzdial.pdf}
\bibfield{author}{\bibinfo{person}{Mark Guzdial}.}
  \bibinfo{year}{2013}\natexlab{}.
\newblock \showarticletitle{Human-Centered Computing: A New Degree for
  Licklider's World}.
\newblock \bibinfo{journal}{\emph{Commun. ACM}} \bibinfo{volume}{56},
  \bibinfo{number}{5} (\bibinfo{date}{may} \bibinfo{year}{2013}),
  \bibinfo{pages}{32–34}.
\newblock
\showISSN{0001-0782}
\urldef\tempurl%
\url{https://doi.org/10.1145/2447976.2447987}
\showDOI{\tempurl}


\bibitem[Ho et~al\mbox{.}(2022)]%
        {imagen-video.pdf}
\bibfield{author}{\bibinfo{person}{Jonathan Ho}, \bibinfo{person}{William
  Chan}, \bibinfo{person}{Chitwan Saharia}, \bibinfo{person}{Jay Whang},
  \bibinfo{person}{Ruiqi Gao}, \bibinfo{person}{Alexey Gritsenko},
  \bibinfo{person}{Diederik~P. Kingma}, \bibinfo{person}{Ben Poole},
  \bibinfo{person}{Mohammad Norouzi}, \bibinfo{person}{David~J. Fleet}, {and}
  \bibinfo{person}{Tim Salimans}.} \bibinfo{year}{2022}\natexlab{}.
\newblock \showarticletitle{Imagen Video: High Definition Video Generation with
  Diffusion Models}.
\newblock  (\bibinfo{year}{2022}).
\newblock
\newblock
\shownote{[Preprint]. Available at: https://arxiv.org/abs/2210.02303 [Accessed
  Nov. 14, 2022].}.


\bibitem[Hoare et~al\mbox{.}(2014)]%
        {2556288.2557298.pdf}
\bibfield{author}{\bibinfo{person}{Michaela Hoare}, \bibinfo{person}{Steve
  Benford}, \bibinfo{person}{Rachel Jones}, {and} \bibinfo{person}{Natasa
  Milic-Frayling}.} \bibinfo{year}{2014}\natexlab{}.
\newblock \showarticletitle{Coming in from the Margins: Amateur Musicians in
  the Online Age}. In \bibinfo{booktitle}{\emph{Proceedings of the SIGCHI
  Conference on Human Factors in Computing Systems}}
  \emph{(\bibinfo{series}{CHI '14})}. \bibinfo{publisher}{Association for
  Computing Machinery}, \bibinfo{address}{New York, NY},
  \bibinfo{pages}{1295–1304}.
\newblock
\showISBNx{9781450324731}
\urldef\tempurl%
\url{https://doi.org/10.1145/2556288.2557298}
\showDOI{\tempurl}


\bibitem[Holtzman et~al\mbox{.}(2020)]%
        {1904.09751.pdf}
\bibfield{author}{\bibinfo{person}{Ari Holtzman}, \bibinfo{person}{Jan Buys},
  \bibinfo{person}{Li Du}, \bibinfo{person}{Maxwell Forbes}, {and}
  \bibinfo{person}{Yejin Choi}.} \bibinfo{year}{2020}\natexlab{}.
\newblock \showarticletitle{The curious case of neural text degeneration}. In
  \bibinfo{booktitle}{\emph{Proceedings of the International Conference on
  Learning Representations}} \emph{(\bibinfo{series}{ICLR '20})}.
  \bibinfo{numpages}{16}~pages.
\newblock


\bibitem[Hong et~al\mbox{.}(2022)]%
        {2205.15868.pdf}
\bibfield{author}{\bibinfo{person}{Wenyi Hong}, \bibinfo{person}{Ming Ding},
  \bibinfo{person}{Wendi Zheng}, \bibinfo{person}{Xinghan Liu}, {and}
  \bibinfo{person}{Jie Tang}.} \bibinfo{year}{2022}\natexlab{}.
\newblock \showarticletitle{CogVideo: Large-scale Pretraining for Text-to-Video
  Generation via Transformers}.
\newblock  (\bibinfo{year}{2022}).
\newblock
\urldef\tempurl%
\url{https://doi.org/10.48550/ARXIV.2205.15868}
\showDOI{\tempurl}
\newblock
\shownote{[Preprint]. Available at: https://arxiv.org/pdf/2205.15868v1.pdf
  [Accessed Nov. 9, 2022].}.


\bibitem[Hutson(2022)]%
        {d41586-022-03479-w.pdf}
\bibfield{author}{\bibinfo{person}{Matthew Hutson}.}
  \bibinfo{year}{2022}\natexlab{}.
\newblock \showarticletitle{Could AI help you to write your next paper?}
\newblock \bibinfo{journal}{\emph{Nature}}  \bibinfo{volume}{611}
  (\bibinfo{year}{2022}), \bibinfo{pages}{192--193}.
\newblock


\bibitem[{Lexica.art}(2022)]%
        {lexica.art}
\bibfield{author}{\bibinfo{person}{{Lexica.art}}.}
  \bibinfo{year}{2022}\natexlab{}.
\newblock \bibinfo{title}{Lexica.art}.
\newblock
\newblock
\newblock
\shownote{{https}://lexica.art/ [Accessed Nov. 9, 2022]}.


\bibitem[Liu and Chilton(2022)]%
        {chilton}
\bibfield{author}{\bibinfo{person}{Vivian Liu} {and} \bibinfo{person}{Lydia~B
  Chilton}.} \bibinfo{year}{2022}\natexlab{}.
\newblock \showarticletitle{Design Guidelines for Prompt Engineering
  Text-to-Image Generative Models}. In \bibinfo{booktitle}{\emph{Proceedings of
  the 2022 CHI Conference on Human Factors in Computing Systems}}
  \emph{(\bibinfo{series}{CHI '22})}. \bibinfo{publisher}{Association for
  Computing Machinery}, \bibinfo{address}{New York, NY}, Article
  \bibinfo{articleno}{384}, \bibinfo{numpages}{23}~pages.
\newblock
\showISBNx{9781450391573}
\urldef\tempurl%
\url{https://doi.org/10.1145/3491102.3501825}
\showDOI{\tempurl}


\bibitem[McCormack et~al\mbox{.}(2023)]%
        {2301.13049.pdf}
\bibfield{author}{\bibinfo{person}{Jon McCormack}, \bibinfo{person}{Camilo~Cruz
  Gambardella}, \bibinfo{person}{Nina Rajcic}, \bibinfo{person}{Stephen~James
  Krol}, \bibinfo{person}{Maria~Teresa Llano}, {and} \bibinfo{person}{Meng
  Yang}.} \bibinfo{year}{2023}\natexlab{}.
\newblock \bibinfo{title}{Is Writing Prompts Really Making Art?}
\newblock
\newblock
\urldef\tempurl%
\url{https://doi.org/10.48550/ARXIV.2301.13049}
\showDOI{\tempurl}


\bibitem[McCormack et~al\mbox{.}(2019)]%
        {aiart}
\bibfield{author}{\bibinfo{person}{Jon McCormack}, \bibinfo{person}{Toby
  Gifford}, {and} \bibinfo{person}{Patrick Hutchings}.}
  \bibinfo{year}{2019}\natexlab{}.
\newblock \showarticletitle{Autonomy, Authenticity, Authorship and Intention in
  Computer Generated Art}. In \bibinfo{booktitle}{\emph{Computational
  Intelligence in Music, Sound, Art and Design}},
  \bibfield{editor}{\bibinfo{person}{Anik{\'o} Ek{\'a}rt},
  \bibinfo{person}{Antonios Liapis}, {and} \bibinfo{person}{Mar{\'i}a~Luz
  Castro~Pena}} (Eds.). \bibinfo{publisher}{Springer International Publishing},
  \bibinfo{address}{Cham}, \bibinfo{pages}{35--50}.
\newblock
\showISBNx{978-3-030-16667-0}


\bibitem[Mialon et~al\mbox{.}(2023)]%
        {2302.07842.pdf}
\bibfield{author}{\bibinfo{person}{Grégoire Mialon}, \bibinfo{person}{Roberto
  Dessì}, \bibinfo{person}{Maria Lomeli}, \bibinfo{person}{Christoforos
  Nalmpantis}, \bibinfo{person}{Ram Pasunuru}, \bibinfo{person}{Roberta
  Raileanu}, \bibinfo{person}{Baptiste Rozière}, \bibinfo{person}{Timo
  Schick}, \bibinfo{person}{Jane Dwivedi-Yu}, \bibinfo{person}{Asli
  Celikyilmaz}, \bibinfo{person}{Edouard Grave}, \bibinfo{person}{Yann LeCun},
  {and} \bibinfo{person}{Thomas Scialom}.} \bibinfo{year}{2023}\natexlab{}.
\newblock \bibinfo{title}{Augmented Language Models: a Survey}.
\newblock
\newblock
\urldef\tempurl%
\url{https://doi.org/10.48550/arXiv.2302.07842}
\showDOI{\tempurl}
\showeprint[arxiv]{2302.07842}~[cs.CL]


\bibitem[Molino and R\'{e}(2021)]%
        {3475167.pdf}
\bibfield{author}{\bibinfo{person}{Piero Molino} {and}
  \bibinfo{person}{Christopher R\'{e}}.} \bibinfo{year}{2021}\natexlab{}.
\newblock \showarticletitle{Declarative Machine Learning Systems}.
\newblock \bibinfo{journal}{\emph{Commun. ACM}} \bibinfo{volume}{65},
  \bibinfo{number}{1} (\bibinfo{date}{dec} \bibinfo{year}{2021}),
  \bibinfo{pages}{42–49}.
\newblock
\showISSN{0001-0782}
\urldef\tempurl%
\url{https://doi.org/10.1145/3475167}
\showDOI{\tempurl}


\bibitem[Neustaedter and Sengers(2012)]%
        {p514-neustaedter.pdf}
\bibfield{author}{\bibinfo{person}{Carman Neustaedter} {and}
  \bibinfo{person}{Phoebe Sengers}.} \bibinfo{year}{2012}\natexlab{}.
\newblock \showarticletitle{Autobiographical Design in HCI Research: Designing
  and Learning through Use-It-Yourself}. In
  \bibinfo{booktitle}{\emph{Proceedings of the Designing Interactive Systems
  Conference}} \emph{(\bibinfo{series}{DIS '12})}.
  \bibinfo{publisher}{Association for Computing Machinery},
  \bibinfo{address}{New York, NY}, \bibinfo{pages}{514–523}.
\newblock
\showISBNx{9781450312103}
\urldef\tempurl%
\url{https://doi.org/10.1145/2317956.2318034}
\showDOI{\tempurl}


\bibitem[{OpenAI}(2022)]%
        {dalle2bias}
\bibfield{author}{\bibinfo{person}{{OpenAI}}.} \bibinfo{year}{2022}\natexlab{}.
\newblock \showarticletitle{Reducing Bias and Improving Safety in DALL·E 2}.
\newblock  (\bibinfo{date}{18 July} \bibinfo{year}{2022}).
\newblock
\newblock
\shownote{{https}://openai.com/blog/reducing-bias-and-improving-safety-in-dall-e-2/
  [Accessed Nov. 9, 2022]}.


\bibitem[{OpenAI}(nd)]%
        {openaidocs}
\bibfield{author}{\bibinfo{person}{{OpenAI}}.} \bibinfo{year}{nd.}\natexlab{}.
\newblock \showarticletitle{Completion -- {OpenAI} {API}.}
\newblock  (\bibinfo{year}{nd.}).
\newblock
\newblock
\shownote{{https}://beta.openai.com/docs/guides/completion [Accessed Nov. 9,
  2022]}.


\bibitem[{OpenArt.ai}(2022)]%
        {openArt.ai}
\bibfield{author}{\bibinfo{person}{{OpenArt.ai}}.}
  \bibinfo{year}{2022}\natexlab{}.
\newblock \bibinfo{title}{OpenArt.ai}.
\newblock
\newblock
\newblock
\shownote{{https}://openart.ai/ [Accessed Nov. 9, 2022]}.


\bibitem[Oppenlaender(2022)]%
        {creativitypromptengineering}
\bibfield{author}{\bibinfo{person}{Jonas Oppenlaender}.}
  \bibinfo{year}{2022}\natexlab{}.
\newblock \showarticletitle{The Creativity of Text-to-Image Generation}. In
  \bibinfo{booktitle}{\emph{Proceedings of the 25th International Academic
  Mindtrek conference}} \emph{(\bibinfo{series}{Academic Mindtrek '22})}.
  \bibinfo{publisher}{ACM}, \bibinfo{numpages}{11 pages}~pages.
\newblock
\urldef\tempurl%
\url{https://doi.org/10.1145/3569219.3569352}
\showDOI{\tempurl}


\bibitem[Parsons(2022)]%
        {dallepromptbook}
\bibfield{author}{\bibinfo{person}{Guy Parsons}.}
  \bibinfo{year}{2022}\natexlab{}.
\newblock \bibinfo{booktitle}{\emph{The DALL·E 2 Prompt Book}}.
\newblock
\newblock
\shownote{{https}://dallery.gallery/wp-content/uploads/2022/07/The-DALL\%C2\%B7E-2-prompt-book-v1.01.pdf
  [Accessed Nov. 9, 2022]}.


\bibitem[Pavlichenko and Ustalov(2022)]%
        {2209.11711.pdf}
\bibfield{author}{\bibinfo{person}{Nikita Pavlichenko} {and}
  \bibinfo{person}{Dmitry Ustalov}.} \bibinfo{year}{2022}\natexlab{}.
\newblock \showarticletitle{Best Prompts for Text-to-Image Models and How to
  Find Them}.
\newblock  (\bibinfo{year}{2022}).
\newblock
\urldef\tempurl%
\url{https://doi.org/10.48550/ARXIV.2209.11711}
\showDOI{\tempurl}
\newblock
\shownote{[Preprint]. Available at: https://arxiv.org/abs/2209.11711 [Accessed
  Nov. 9, 2022].}.


\bibitem[Pink et~al\mbox{.}(2016)]%
        {pink2016}
\bibfield{author}{\bibinfo{person}{Sarah Pink}, \bibinfo{person}{Heather
  Horst}, \bibinfo{person}{John Postill}, \bibinfo{person}{Larissa Hjorth},
  \bibinfo{person}{Tania Lewis}, {and} \bibinfo{person}{Jo Tacchi}.}
  \bibinfo{year}{2016}\natexlab{}.
\newblock \bibinfo{booktitle}{\emph{Digital Ethnography: Principles and
  Practice}}.
\newblock \bibinfo{publisher}{SAGE}, \bibinfo{address}{London, UK}.
\newblock


\bibitem[Postill and Pink(2012)]%
        {postill2012.pdf}
\bibfield{author}{\bibinfo{person}{John Postill} {and} \bibinfo{person}{Sarah
  Pink}.} \bibinfo{year}{2012}\natexlab{}.
\newblock \showarticletitle{Social Media Ethnography: The Digital Researcher in
  a Messy Web}.
\newblock \bibinfo{journal}{\emph{Media International Australia}}
  \bibinfo{volume}{145}, \bibinfo{number}{1} (\bibinfo{year}{2012}),
  \bibinfo{pages}{123--134}.
\newblock
\urldef\tempurl%
\url{https://doi.org/10.1177/1329878X1214500114}
\showDOI{\tempurl}


\bibitem[Pressman et~al\mbox{.}(2022)]%
        {pressmancrowson2022}
\bibfield{author}{\bibinfo{person}{John~David Pressman},
  \bibinfo{person}{Katherine Crowson}, {and}
  \bibinfo{person}{Simulacra~Captions Contributors}.}
  \bibinfo{year}{2022}\natexlab{}.
\newblock \bibinfo{booktitle}{\emph{Simulacra Aesthetic Captions}}.
\newblock \bibinfo{type}{{T}echnical {R}eport} Version 1.0.
  \bibinfo{institution}{Stability AI}.
\newblock
\newblock
\shownote{\ url { https://github.com/JD-P/simulacra-aesthetic-captions }}.


\bibitem[Qiao et~al\mbox{.}(2022)]%
        {3527927.3532792.pdf}
\bibfield{author}{\bibinfo{person}{Han Qiao}, \bibinfo{person}{Vivian Liu},
  {and} \bibinfo{person}{Lydia Chilton}.} \bibinfo{year}{2022}\natexlab{}.
\newblock \showarticletitle{Initial Images: Using Image Prompts to Improve
  Subject Representation in Multimodal AI Generated Art}. In
  \bibinfo{booktitle}{\emph{Creativity and Cognition}}
  \emph{(\bibinfo{series}{C\&C '22})}. \bibinfo{publisher}{Association for
  Computing Machinery}, \bibinfo{address}{New York, NY},
  \bibinfo{pages}{15–28}.
\newblock
\showISBNx{9781450393270}
\urldef\tempurl%
\url{https://doi.org/10.1145/3527927.3532792}
\showDOI{\tempurl}


\bibitem[Radford et~al\mbox{.}(2021)]%
        {CLIP}
\bibfield{author}{\bibinfo{person}{Alec Radford}, \bibinfo{person}{Jong~Wook
  Kim}, \bibinfo{person}{Chris Hallacy}, \bibinfo{person}{Aditya Ramesh},
  \bibinfo{person}{Gabriel Goh}, \bibinfo{person}{Sandhini Agarwal},
  \bibinfo{person}{Girish Sastry}, \bibinfo{person}{Amanda Askell},
  \bibinfo{person}{Pamela Mishkin}, \bibinfo{person}{Jack Clark},
  \bibinfo{person}{Gretchen Krueger}, {and} \bibinfo{person}{Ilya Sutskever}.}
  \bibinfo{year}{2021}\natexlab{}.
\newblock \showarticletitle{Learning Transferable Visual Models From Natural
  Language Supervision}. In \bibinfo{booktitle}{\emph{Proceedings of the 38th
  International Conference on Machine Learning}}
  \emph{(\bibinfo{series}{Proceedings of Machine Learning Research},
  Vol.~\bibinfo{volume}{139})}, \bibfield{editor}{\bibinfo{person}{Marina
  Meila} {and} \bibinfo{person}{Tong Zhang}} (Eds.). \bibinfo{publisher}{PMLR},
  \bibinfo{pages}{8748--8763}.
\newblock
\urldef\tempurl%
\url{https://proceedings.mlr.press/v139/radford21a.html}
\showURL{%
\tempurl}


\bibitem[Ramchurn et~al\mbox{.}(2021)]%
        {RAMCHURN2021102891}
\bibfield{author}{\bibinfo{person}{Sarvapali~D. Ramchurn},
  \bibinfo{person}{Sebastian Stein}, {and} \bibinfo{person}{Nicholas~R.
  Jennings}.} \bibinfo{year}{2021}\natexlab{}.
\newblock \showarticletitle{Trustworthy human-{AI} partnerships}.
\newblock \bibinfo{journal}{\emph{iScience}} \bibinfo{volume}{24},
  \bibinfo{number}{8} (\bibinfo{year}{2021}), \bibinfo{numpages}{13}~pages.
\newblock
\urldef\tempurl%
\url{https://doi.org/10.1016/j.isci.2021.102891}
\showDOI{\tempurl}


\bibitem[Ramesh et~al\mbox{.}(2021)]%
        {2102.12092.pdf}
\bibfield{author}{\bibinfo{person}{Aditya Ramesh}, \bibinfo{person}{Mikhail
  Pavlov}, \bibinfo{person}{Gabriel Goh}, \bibinfo{person}{Scott Gray},
  \bibinfo{person}{Chelsea Voss}, \bibinfo{person}{Alec Radford},
  \bibinfo{person}{Mark Chen}, {and} \bibinfo{person}{Ilya Sutskever}.}
  \bibinfo{year}{2021}\natexlab{}.
\newblock \showarticletitle{Zero-Shot Text-to-Image Generation}. In
  \bibinfo{booktitle}{\emph{Proceedings of the 38th International Conference on
  Machine Learning}} \emph{(\bibinfo{series}{Proceedings of Machine Learning
  Research}, Vol.~\bibinfo{volume}{139})},
  \bibfield{editor}{\bibinfo{person}{Marina Meila} {and} \bibinfo{person}{Tong
  Zhang}} (Eds.). \bibinfo{publisher}{PMLR}, \bibinfo{pages}{8821--8831}.
\newblock


\bibitem[Reynolds and McDonell(2021)]%
        {2102.07350.pdf}
\bibfield{author}{\bibinfo{person}{Laria Reynolds} {and} \bibinfo{person}{Kyle
  McDonell}.} \bibinfo{year}{2021}\natexlab{}.
\newblock \showarticletitle{Prompt Programming for Large Language Models:
  Beyond the Few-Shot Paradigm}. In \bibinfo{booktitle}{\emph{Extended
  Abstracts of the 2021 CHI Conference on Human Factors in Computing Systems}}
  \emph{(\bibinfo{series}{CHI EA '21})}. \bibinfo{publisher}{Association for
  Computing Machinery}, \bibinfo{address}{New York, NY}, Article
  \bibinfo{articleno}{314}, \bibinfo{numpages}{7}~pages.
\newblock
\showISBNx{9781450380959}
\urldef\tempurl%
\url{https://doi.org/10.1145/3411763.3451760}
\showDOI{\tempurl}


\bibitem[Richards(2023)]%
        {AutoGPT}
\bibfield{author}{\bibinfo{person}{Toran~Bruce Richards}.}
  \bibinfo{year}{2023}\natexlab{}.
\newblock \bibinfo{title}{Significant-Gravitas/Auto-GPT GitHub repository}.
\newblock
  \bibinfo{howpublished}{\url{https://github.com/Significant-Gravitas/Auto-GPT}}.
\newblock


\bibitem[Rombach et~al\mbox{.}(2021)]%
        {latent-diffusion}
\bibfield{author}{\bibinfo{person}{Robin Rombach}, \bibinfo{person}{Andreas
  Blattmann}, \bibinfo{person}{Dominik Lorenz}, \bibinfo{person}{Patrick
  Esser}, {and} \bibinfo{person}{Björn Ommer}.}
  \bibinfo{year}{2021}\natexlab{}.
\newblock \showarticletitle{High-Resolution Image Synthesis with Latent
  Diffusion Models.}
\newblock  (\bibinfo{year}{2021}).
\newblock
\showeprint[arxiv]{2112.10752}~[cs.CV]
\newblock
\shownote{[Preprint]. Available at: https://arxiv.org/abs/2112.10752 [Accessed
  Nov. 9, 2022].}.


\bibitem[Rombach et~al\mbox{.}(2022)]%
        {2207.13038.pdf}
\bibfield{author}{\bibinfo{person}{Robin Rombach}, \bibinfo{person}{Andreas
  Blattmann}, {and} \bibinfo{person}{Björn Ommer}.}
  \bibinfo{year}{2022}\natexlab{}.
\newblock \showarticletitle{Text-Guided Synthesis of Artistic Images with
  Retrieval-Augmented Diffusion Models}.
\newblock  (\bibinfo{year}{2022}).
\newblock
\newblock
\shownote{[Preprint]. Available at: https://arxiv.org/abs/2207.13038 [Accessed
  Nov. 9, 2022].}.


\bibitem[Schuhmann(2022)]%
        {laion-aesthetics}
\bibfield{author}{\bibinfo{person}{Christoph Schuhmann}.}
  \bibinfo{year}{2022}\natexlab{}.
\newblock \bibinfo{title}{LAION-Aesthetics}.
\newblock
\newblock
\urldef\tempurl%
\url{https://laion.ai/blog/laion-aesthetics/}
\showURL{%
\tempurl}
\newblock
\shownote{{https}://laion.ai/blog/laion-aesthetics/ [Accessed Nov. 11, 2022]}.


\bibitem[Shneiderman(2020)]%
        {2002.04087.pdf}
\bibfield{author}{\bibinfo{person}{Ben Shneiderman}.}
  \bibinfo{year}{2020}\natexlab{}.
\newblock \showarticletitle{Human-Centered Artificial Intelligence: Reliable,
  Safe \& Trustworthy}.
\newblock \bibinfo{journal}{\emph{International Journal of Human–Computer
  Interaction}} \bibinfo{volume}{36}, \bibinfo{number}{6}
  (\bibinfo{year}{2020}), \bibinfo{pages}{495--504}.
\newblock
\urldef\tempurl%
\url{https://doi.org/10.1080/10447318.2020.1741118}
\showDOI{\tempurl}


\bibitem[Singer et~al\mbox{.}(2022)]%
        {2209.14792.pdf}
\bibfield{author}{\bibinfo{person}{Uriel Singer}, \bibinfo{person}{Adam
  Polyak}, \bibinfo{person}{Thomas Hayes}, \bibinfo{person}{Xi Yin},
  \bibinfo{person}{Jie An}, \bibinfo{person}{Songyang Zhang},
  \bibinfo{person}{Qiyuan Hu}, \bibinfo{person}{Harry Yang},
  \bibinfo{person}{Oron Ashual}, \bibinfo{person}{Oran Gafni},
  \bibinfo{person}{Devi Parikh}, \bibinfo{person}{Sonal Gupta}, {and}
  \bibinfo{person}{Yaniv Taigman}.} \bibinfo{year}{2022}\natexlab{}.
\newblock \showarticletitle{Make-A-Video: Text-to-Video Generation without
  Text-Video Data}.
\newblock  (\bibinfo{year}{2022}).
\newblock
\urldef\tempurl%
\url{https://doi.org/10.48550/ARXIV.2209.14792}
\showDOI{\tempurl}
\newblock
\shownote{[Preprint]. Available at: https://arxiv.org/abs/2209.14792 [Accessed
  Nov. 14, 2022].}.


\bibitem[Smith(2022)]%
        {travelersguide}
\bibfield{author}{\bibinfo{person}{Ethan Smith}.}
  \bibinfo{year}{2022}\natexlab{}.
\newblock \showarticletitle{A Traveler's Guide to the Latent Space.}
\newblock  (\bibinfo{year}{2022}).
\newblock
\newblock
\shownote{{https}://sweet-hall-e72.notion.site/A-Traveler-s-Guide-to-the-Latent-Space-85efba7e5e6a40e5bd3cae980f30235f
  [Accessed Nov. 9, 2022]}.


\bibitem[Snell(2021)]%
        {aliendreams}
\bibfield{author}{\bibinfo{person}{Charlie Snell}.}
  \bibinfo{year}{2021}\natexlab{}.
\newblock \showarticletitle{Alien Dreams: An Emerging Art Scene.}
\newblock  (\bibinfo{year}{2021}).
\newblock
\newblock
\shownote{{https}://ml.berkeley.edu/blog/posts/clip-art/ [Accessed Nov. 9,
  2022]}.


\bibitem[Villegas et~al\mbox{.}(2022)]%
        {phenaki.pdf}
\bibfield{author}{\bibinfo{person}{Ruben Villegas}, \bibinfo{person}{Mohammad
  Babaeizadeh}, \bibinfo{person}{Pieter-Jan Kindermans},
  \bibinfo{person}{Hernan Moraldo}, \bibinfo{person}{Han Zhang},
  \bibinfo{person}{Mohammad~Taghi Saffar}, \bibinfo{person}{Santiago Castro},
  \bibinfo{person}{Julius Kunze}, {and} \bibinfo{person}{Dumitru Erhan}.}
  \bibinfo{year}{2022}\natexlab{}.
\newblock \showarticletitle{Phenaki: Variable Length Video Generation from Open
  Domain Textual Descriptions}.
\newblock  (\bibinfo{year}{2022}).
\newblock
\newblock
\shownote{{https}://openreview.net/forum?id=vOEXS39nOF [Accessed Nov. 14,
  2022]}.


\bibitem[Wang et~al\mbox{.}(2022)]%
        {2210.14896.pdf}
\bibfield{author}{\bibinfo{person}{Zijie~J. Wang}, \bibinfo{person}{Evan
  Montoya}, \bibinfo{person}{David Munechika}, \bibinfo{person}{Haoyang Yang},
  \bibinfo{person}{Benjamin Hoover}, {and} \bibinfo{person}{Duen~Horng Chau}.}
  \bibinfo{year}{2022}\natexlab{}.
\newblock \showarticletitle{DiffusionDB: A Large-scale Prompt Gallery Dataset
  for Text-to-Image Generative Models}.
\newblock  (\bibinfo{year}{2022}).
\newblock
\urldef\tempurl%
\url{https://doi.org/10.48550/ARXIV.2210.14896}
\showDOI{\tempurl}
\newblock
\shownote{[Preprint]. Available at: https://arxiv.org/abs/2210.14896 [Accessed
  Nov. 9, 2022].}.


\bibitem[Wobbrock and Kientz(2016)]%
        {wobbrock}
\bibfield{author}{\bibinfo{person}{Jacob~O. Wobbrock} {and}
  \bibinfo{person}{Julie~A. Kientz}.} \bibinfo{year}{2016}\natexlab{}.
\newblock \showarticletitle{Research Contributions in Human-Computer
  Interaction}.
\newblock \bibinfo{journal}{\emph{Interactions}} \bibinfo{volume}{23},
  \bibinfo{number}{3} (\bibinfo{year}{2016}), \bibinfo{pages}{38–44}.
\newblock
\showISSN{1072-5520}
\urldef\tempurl%
\url{https://doi.org/10.1145/2907069}
\showDOI{\tempurl}


\bibitem[Zaremba and Brockman(2021)]%
        {codex}
\bibfield{author}{\bibinfo{person}{Wojciech Zaremba} {and}
  \bibinfo{person}{Greg Brockman}.} \bibinfo{year}{2021}\natexlab{}.
\newblock \showarticletitle{OpenAI Codex.}
\newblock  (\bibinfo{year}{2021}).
\newblock
\newblock
\shownote{{https}://openai.com/blog/openai-codex [Accessed Nov. 9, 2022]}.


\bibitem[Zhang et~al\mbox{.}(2020)]%
        {3394171.3414017.pdf}
\bibfield{author}{\bibinfo{person}{Lisai Zhang}, \bibinfo{person}{Qingcai
  Chen}, \bibinfo{person}{Baotian Hu}, {and} \bibinfo{person}{Shuoran Jiang}.}
  \bibinfo{year}{2020}\natexlab{}.
\newblock \bibinfo{booktitle}{\emph{Text-Guided Neural Image Inpainting}}.
\newblock \bibinfo{publisher}{Association for Computing Machinery},
  \bibinfo{address}{New York, NY}, \bibinfo{pages}{1302–1310}.
\newblock
\showISBNx{9781450379885}
\urldef\tempurl%
\url{https://doi.org/10.1145/3394171.3414017}
\showDOI{\tempurl}


\end{thebibliography}

\end{document}